\documentclass[aps,prd,reprint,superscriptaddress,nofootinbib,amsmath,amssymb,longbibliography,floatfix]{revtex4-2}
\usepackage[T1]{fontenc}
\usepackage{lmodern}
\usepackage{array}[=2016-10-06]
\usepackage{graphicx,booktabs,tabularx}
\usepackage{bm}
\usepackage[unicode,breaklinks=true,colorlinks=true,linkcolor=blue,citecolor=blue,urlcolor=blue]{hyperref}
\usepackage{orcidlink}
\newcolumntype{L}{>{\raggedright\arraybackslash}X}
\newcolumntype{C}{>{\centering\arraybackslash}X}
\newcolumntype{R}{>{\raggedleft\arraybackslash}X}
\begin{document}

\title{Flavor decoherence and washout timing in nonthermal leptogenesis after inflation}

\author{Boliang Yu\,\orcidlink{0009-0006-4664-5068}}
\email[Contact author: ]{yuboliang@sjtu.edu.cn}
\affiliation{School of Physics and Astronomy, Shanghai Jiao Tong University, 800 Dongchuan Road, Shanghai 200240, China}
\author{Ruixin Zhou\,\orcidlink{0009-0004-9303-6674}}
\affiliation{School of Physics and Astronomy, Shanghai Jiao Tong University, 800 Dongchuan Road, Shanghai 200240, China}

\begin{abstract}

We investigate why changing the heavy-neutrino mass hierarchy can reverse the baryon asymmetry in nonthermal leptogenesis after hilltop inflation. Along the mass trajectory studied here, lowering the mass of the lighter state, \(N_1\), reverses the final asymmetry while leaving the dominant injection from \(N_2\) decays nearly unchanged. The mass change primarily delays \(N_1\) washout relative to electron--muon flavor decoherence; at fixed source and background, translating the washout history while preserving its shape and integrated strength reproduces more than 99\% of the yield changes from the nonthermal \(N_2\) source at two physical target masses. Decomposing the \(N_2\) CP source shows that the projector-aligned component remains positive under both washout histories, whereas the initially trace-free component reverses its propagated contribution and accounts for about 99.7\% of the timing contrast in the fixed-source comparison. Rapid damping of \(e\mu\) flavor coherence largely removes this contrast, while rapid chemical equilibration does not. Later washout therefore acts on a more decohered flavor state, reducing the positive \(e\mu\) contribution and allowing negative diagonal contributions to determine the final sign. At a mass recomputed to reproduce the observed baryon yield, a 10\% increase of the muon collision operator reverses the sign in a fixed-input sensitivity test.

\end{abstract}

\maketitle

\section{Introduction}

Nonthermal leptogenesis converts the flavor charge produced by heavy-neutrino decay into a net baryon asymmetry. Heavy Majorana-neutrino decays realize out-of-equilibrium baryogenesis with lepton-number and CP violation \cite{Y10_sakharov,Y10_fukugita}, and anomalous electroweak processes redistribute baryon and lepton number \cite{Y10_kuzmin}. This connects neutrino mass to the observed matter asymmetry \cite{Y10_davidson}. Production of the heavy states by inflaton decay gives the nonthermal setting \cite{Y10_asaka_decay,Y10_asaka_inflation}. Here \(N_1\) and \(N_2\) are heavy Majorana neutrinos ordered by mass. Nonthermal \(N_2\) decays provide the dominant asymmetry, which is subsequently processed by inverse decay and scattering involving the lighter \(N_1\).

Lepton flavor changes both CP production and washout, as established by early flavor-resolved studies \cite{Y10_barbieri,Y10_abada_issues,Y10_nardi_flavor} and subsequent quantitative analyses \cite{Y10_abada_matters,Y10_blanchet_flavor}. Spectator effects change the distribution of an asymmetry among particle species while conserving baryon-minus-lepton asymmetry directly \cite{Y10_spectator,Y10_higgs_sphaleron}. Such redistribution can nevertheless change the amount that survives later lepton-number-violating scattering or inverse decay. The equilibrium spectator construction and its partially equilibrated extension \cite{Y10_old04} supply the chemical setting. An asymmetry generated by \(N_2\) can contain components protected from subsequent \(N_1\) washout \cite{Y10_n2}, and flavor coupling modifies that survival \cite{Y10_old05}. Density-matrix evolution describes the gradual loss of flavor coherence between limiting flavor regimes and the trace-free flavor information associated with phantom terms \cite{Y10_old03,Y10_old06,Y10_old07}. The relevant competition involves flavor-sensitive interactions and inverse decay, as well as expansion \cite{Y10_zeno}. Flavor-induced sign reversal, including transitions involving the muon Yukawa interaction, has already been demonstrated \cite{Y10_old08}. Non-instantaneous reheating further changes the relation between temperature and charged-lepton equilibration \cite{Y10_old09}. The kinetic descriptions and flavor regimes are reviewed in Ref. \cite{Y10_flavor_review}.

\begin{figure*}[t]

\centering

\includegraphics[width=\textwidth]{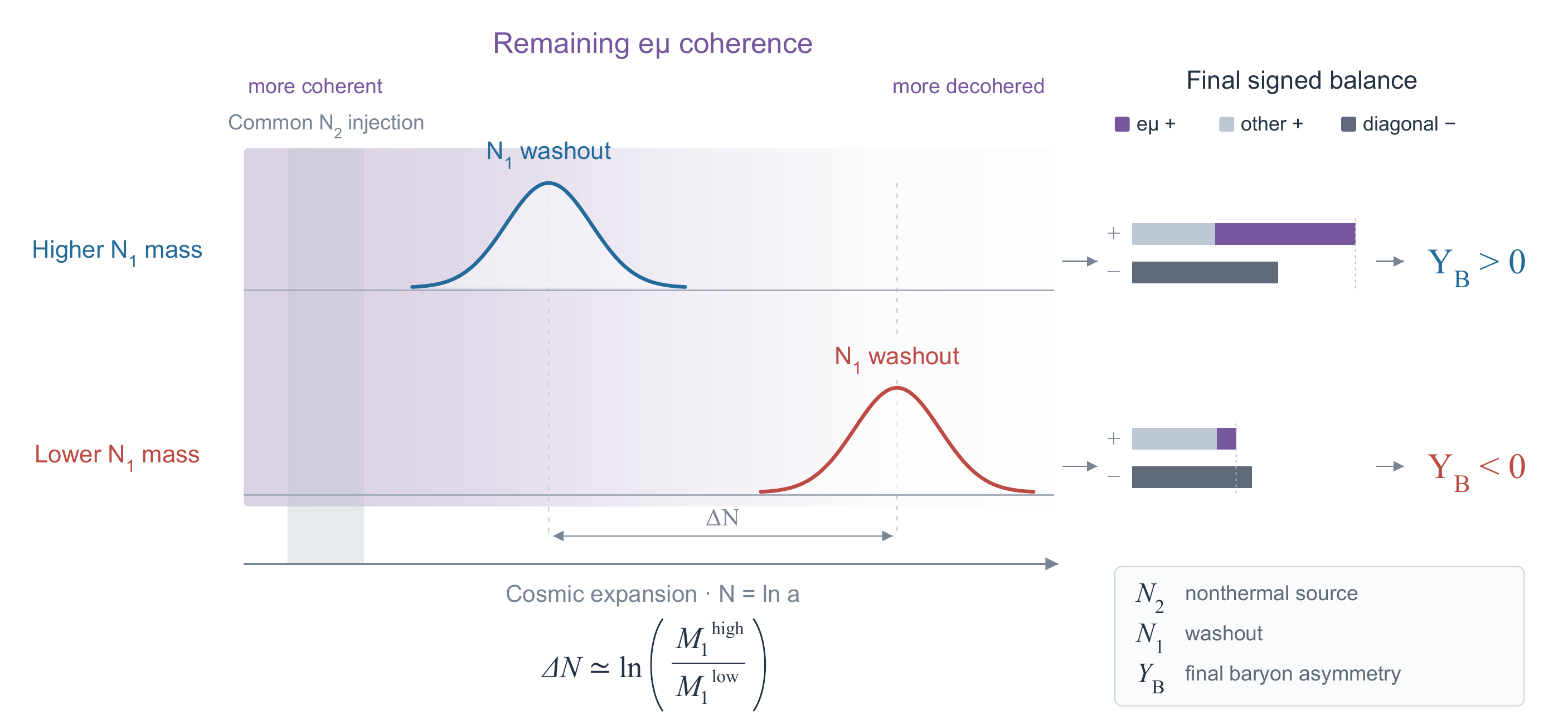}

\caption{Washout timing and the sign of the baryon asymmetry. Nearly unchanged nonthermal \(N_2\) injection is followed by earlier or later \(N_1\) washout as \(e\mu\) flavor decoherence proceeds. The shared expansion coordinate makes the mass-dependent delay visible. Purple shading and the injection window are schematic. The signed bars summarize the original and translated budgets for the nonthermal \(N_2\) source in Table~\ref{tab:6}: the positive \(e\mu\) contribution decreases, leaving a negative total balance. Light gray collects the remaining net positive contribution, including injection; dark gray shows the negative lepton-diagonal terms. \(N_1\) and \(N_2\) are heavy Majorana neutrinos.}\label{fig:1}

\end{figure*}

In our calculation, inflaton decay supplies the nonthermal population, whose subsequent decays inject flavor charge. The charge injected by \(N_2\) contains compensating flavor contributions and off-diagonal correlations, which we follow with a charge-density matrix. Charged-lepton interactions redistribute this charge and cause flavor decoherence while \(N_1\) interactions wash it out in a flavor-dependent way. These processes overlap in time. Selective washout changes the balance of contributions to the total charge, so nearly unchanged injection can produce opposite final signs. Determining which part of this evolution changes the sign is the physical problem addressed here.

We ask why lowering the lightest heavy-neutrino mass reverses the final asymmetry along a trajectory with nearly unchanged nonthermal \(N_2\) injection. The mass change can affect the strength, flavor structure and timing of washout. Its timing dependence has a direct physical basis: on a cooling branch, a smaller \(M_1\), the mass of \(N_1\), reaches a given \(M_1/T\) later. Washout then acts after further flavor decoherence. We determine how much of the sign contrast this delay explains and which components of the injected charge are responsible.

The studied system is a Type-I seesaw realization of hilltop inflation, motivated by Ref. \cite{Y10_old01}. The inflaton potential and inflaton-neutrino coupling determine the production of the heavy states and the reheating history. We retain momentum-dependent nonthermal populations and the Standard Model charge-density matrices, with thermal and nonthermal sources from both \(N_1\) and \(N_2\). Injection samples continuing entropy production, while the delayed washout mainly samples the subsequent radiation era. The mass trajectory holds the potential and complex-rotation parameter fixed. Appendix~\ref{app:A} specifies the model inputs and reheating history used in the transport calculation. The axion sector and the broader cosmological consistency of this realization are outside the scope of this paper.

Lowering \(M_1/M_2\) from \(6.30524\times10^{-3}\) to \(7.5\times10^{-4}\) changes the normalized all-source ratio \(\mathcal{R}_{\rm all}\) from about \(+26.7\) to \(-5.28\), while the integrated norm of the nonthermal \(N_2\) injection changes by only \(3.94\times10^{-5}\) relatively. Translations fixed by the washout profiles reproduce more than 99\% of the abundance changes to two target masses. In the fixed nonthermal \(N_2\) comparison, the initially trace-free source component supplies about 99.7\% of the yield difference between the original and translated histories and makes the delayed result negative. The signed budget shows how this occurs: later washout has a smaller positive contribution from \(e\mu\) charge, and the balance with diagonal contributions becomes negative. Fig.~\ref{fig:1} summarizes this mechanism. At the abundance-matched mass, increasing the muon collision operator by 10\% reverses the sign in a fixed-input sensitivity test using the late Standard Model conversion.

The paper is organized as follows. Section~\ref{sec:mass_clock} presents the physical realization and mass trajectory, and tests how the mass hierarchy changes washout timing. Section~\ref{sec:flavor_charge} connects the source decomposition, flavor decoherence and signed washout contributions to the sign reversal. Section~\ref{sec:baryon_yield} examines the consequences for the baryon yield and its response to charged-lepton collisions. Section~\ref{sec:conclusions} gives our conclusions. The appendices provide the model and collision inputs, details of the controlled tests, numerical checks, and late baryon conversion.

\section{Mass hierarchy as a washout clock}\label{sec:mass_clock}

\subsection{Physical realization}

The link between hilltop inflation and nonthermal leptogenesis is carried by a common inflaton vacuum and decay history. Hilltop potentials provide a small-field realization of slow-roll inflation \cite{Y10_hilltop}. The heavy states have masses \(M_1<M_2<M_3\). Here \(m_\phi\) is the inflaton mass.

At the reference background, the maximum bath temperature is \(T_{\max}\simeq1.0\times10^{11}\ {\rm GeV}\), while \(M_2\simeq6.64\times10^{11}\ {\rm GeV}\) and \(m_\phi/2\simeq1.05\times10^{12}\ {\rm GeV}\). The three displayed masses span \(M_1\simeq5.0\times10^8\)--\(4.2\times10^9\ {\rm GeV}\), giving \(M_1<T_{\max}<M_2<m_\phi/2\). Inflaton decay can therefore produce \(N_2\), while the bath reaches temperatures above \(M_1\). The thermal and nonthermal contributions are calculated from the evolving populations and sources. The main calculation uses the realization near an inflection point, with \(M_2=m_\phi/\sqrt{10}\) and \(M_3/M_2=100\). The mass trajectory changes \(M_1/M_2\), keeping the potential coefficients, complex-rotation parameter \(w\), real rotation angles and low-energy neutrino inputs fixed. Each physical mass has its own recomputed background and Yukawa matrix, obtained with one-loop running and sequential threshold matching \cite{Y10_antusch_operator,Y10_old10}. The detailed inputs are collected in Appendix~\ref{app:A}.

The Yukawa matrix \(Y\) of the heavy states has a light-flavor row and a heavy-state column. Let \(M_R\) be the heavy Majorana mass matrix, \(m_\nu\) the effective light-neutrino mass matrix, and \(v_{\rm EW}=174\ {\rm GeV}\) the Higgs vacuum expectation value in our Yukawa convention. In the Type-I seesaw mechanism \cite{Y10_minkowski,Y10_yanagida,Y10_gellmann,Y10_mohapatra}, eliminating the heavy fields for \(\|v_{\rm EW}YM_R^{-1}\|\ll1\) gives

\begin{equation}
m_\nu=-v_{\rm EW}^{\,2}YM_R^{-1}Y^T .
\label{eq:4}
\end{equation}
For each heavy state, \(Y_i\) is column \(i\) of \(Y\), and \(P_i=Y_iY_i^\dagger/(Y^\dagger Y)_{ii}\) is its normalized flavor projector.

The effective neutrino mass and radiation-reference decay parameter are standard diagnostics of leptogenesis \cite{Y10_davidson,Y10_pedestrians}. We report them in the numerical convention used for the mass trajectory:

\begin{equation}
\begin{gathered}
\widetilde m_i={v_{\rm EW}^2[Y^\dagger Y]_{ii}(M_i)\over M_i},\\
K_i={\widetilde m_i\over m_*},\\
m_*=1.08\times10^{-3}\ {\rm eV}.
\end{gathered}
\label{eq:6}
\end{equation}

\subsection{Charge evolution}

We use \(N=\ln(a/a_{\rm initial})\) for evolution; a prime denotes \(d/dN\). Here \(a\) is the scale factor and \(a_{\rm initial}\) its initial value.

The heavy-neutrino populations obey Boltzmann equations. In a homogeneous and isotropic expanding universe, their conventional distribution-function form is \cite{Y10_hahn_boltzmann}

\begin{equation*}
\left(\frac{\partial}{\partial t}
-Hp\frac{\partial}{\partial p}\right)f_i(t,p)
=\mathcal C_i[f].
\end{equation*}

Here \(f_i\) is the phase-space distribution of heavy-neutrino species \(N_i\), \(t\) is cosmic time, \(p\) is physical momentum, and \(H=\dot a/a\) is the Hubble parameter for scale factor \(a\). The left-hand side describes time evolution and momentum redshift; the collision term \(\mathcal C_i\) gives the interaction-induced change per unit time. Integrating this equation over a comoving momentum bin gives the equation for physical number density used below.

Reheating determines the temperature \(T\), Hubble parameter \(H\), and entropy density \(s_{\rm ent}\) throughout production and decay. During perturbative reheating the maximum temperature and the later radiation-dominated temperature are distinct, and comoving entropy need not yet be conserved \cite{Y10_tmax,Y10_reheat_review}. Integrating the distribution-function Boltzmann equation above over a fixed comoving momentum bin gives the equation for the physical number density of nonthermal heavy-neutrino populations:

\begin{equation*}
\frac{dn_{ig}}{dt}+3Hn_{ig}
=J_{ig}-\Gamma_i^{\rm rest}\frac{M_i}{E_{ig}}n_{ig}.
\end{equation*}

Here \(n_{ig}\) is the physical number density of species \(N_i\) in comoving momentum bin \(g\), \(t\) is cosmic time and \(H=\dot a/a\). Inflaton decay supplies particles at the rate \(J_{ig}\) per physical volume and time, \(3Hn_{ig}\) describes dilution, and \(\Gamma_i^{\rm rest}M_i/E_{ig}\) is the decay rate including time dilation, with rest-frame decay width \(\Gamma_i^{\rm rest}\). Physical momentum redshifts as \(p\propto a^{-1}\), with \(E_{ig}=\sqrt{M_i^2+p_{c,g}^2/a^2}\) and fixed comoving momentum \(p_{c,g}\). These decay losses determine the nonthermal charge source. 

The background contains the inflaton, radiation and nonthermal heavy-neutrino energy densities. Thermal heavy-neutrino populations enter the source and washout, without an additional energy term for thermal heavy particles in the Friedmann equation.

For each Standard Model gauge multiplet \(X=Q,U,D,\ell,E\), we use a Hermitian particle-minus-antiparticle charge-density matrix \(q_X\). The labels denote the quark doublet, up-type quark, down-type quark, lepton doublet and right-handed charged lepton, respectively. We normalize these matrices to the final comoving entropy \(S_f\). With \(n_{\Delta X}\) the corresponding physical charge-density matrix and \(S_c(N)=a^3s_{\rm ent}(N)\),

\begin{equation}
\begin{gathered}
q_X(N)=\frac{a^3 n_{\Delta X}(N)}{S_f},\\
r_s(N)=\frac{S_c(N)}{S_f},\\
q_X=\frac{n_{\Delta X}}{s_{\rm ent}}r_s .
\end{gathered}
\label{eq:3}
\end{equation}

These variables retain dilution while entropy is produced. Inflationary scale matching uses the same reheating and final-entropy history \cite{Y10_liddle}.

The baryon-minus-lepton asymmetry normalized to final entropy is

\begin{equation}
Y_{B-L}=\frac13\operatorname{Tr}(q_Q+q_U+q_D)
-\operatorname{Tr}(q_\ell+q_E).
\label{eq:8}
\end{equation}

The primary output is \(Y_{B-L}\) at the final entropy surface \(N_f\), the end of charge evolution. The calligraphic ratio \(\mathcal{R}_s\) always uses the reference conversion \(c_0\) in transport tables and curves; its hatted counterpart \(\widehat{\mathcal{R}}_s\) uses the computed late-time conversion \(c_{\rm eff}\):

\begin{equation}
\begin{aligned}
\mathcal{R}_s&={c_0Y_{B-L}^{(s)}(N_f)\over Y_B^{\rm obs}},\qquad
\widehat{\mathcal{R}}_s={Y_B^{(s)}\over Y_B^{\rm obs}}
={c_{\rm eff}\over c_0}\mathcal{R}_s,\\
c_0&=0.315,\qquad Y_B^{\rm obs}=8.7\times10^{-11},
\qquad s\in\{{\rm nt2},{\rm all}\}.
\end{aligned}
\label{eq:9}
\end{equation}

The observational reference is a rounded benchmark inferred from the cosmological baryon density \cite{Y10_planck_params}: \(\Omega_bh^2=0.02237\) gives \(Y_B\simeq8.7\times10^{-11}\) after the standard photon-to-entropy conversion. The coefficient \(c_0\) is an adopted normalization also quoted in Ref. \cite{Y10_babu_minimal_so10}; its distinction from the computed late conversion is quantified below. The label \({\rm nt2}\) denotes the nonthermal \(N_2\) source alone; \({\rm all}\) denotes the sum of thermal and nonthermal \(N_1\) and \(N_2\) sources. Linearity allows this decomposition on a fixed background. Thus \(\mathcal{R}_{\rm all}=1\) is a useful reference point, and \(\widehat{\mathcal{R}}_{\rm all}=1\) is the physical abundance target in the adopted conversion approximation.

Once the background and heavy-neutrino population histories have been determined, we collect the real components of the charge and auxiliary matrices in \(\mathbf x\). The source contributions form \(\mathbf b(N)\), while washout, charged-lepton and quark collisions, sphaleron transitions and dispersive couplings form \(A(N)\). For either source choice \(s\), the charge evolution and final observable have the form

\begin{equation}
\mathbf x'=A(N)\mathbf x+\mathbf b(N),\qquad
\mathcal{R}_s=\mathbf c^T\mathbf x(N_f).
\label{eq:10}
\end{equation}

The constant vector \(\mathbf c\) implements Eq.~\eqref{eq:9}. A charge-conserving redistribution affects the final asymmetry through subsequent lepton-number violation. Production, redistribution and washout therefore enter in their time order.

We write \(A=A_{\rm rest}+W_1\). The operator \(W_1\) contains \(N_1\) inverse decay and \(\Delta L=1\) top-quark and gauge scattering, including gain, loss and Higgs-charge terms. The \(\Delta L=2\) operator belongs to \(A_{\rm rest}\).

Four sources arise from thermal and nonthermal \(N_1\) and \(N_2\). The reference timing comparisons use zero thermal, charge and auxiliary initial conditions at the maximum-temperature surface. We also evolve the thermal populations and the full charge and auxiliary state from the first vacuum crossing through the temperature maximum and continue their evolution thereafter. The physical abundance match and mechanism comparisons with alternative charged-lepton collision prescriptions include this continuation. The calculation uses a thermal bath and linear charge evolution; the mass comparisons satisfy \(\max|\mu_{\rm chem}/T|<2.1\times10^{-4}\).

\subsection{Physical mass trajectory}

The self-consistently recomputed mass trajectory gives a large sign contrast: lowering \(M_1/M_2\) from \(6.30524\times10^{-3}\) to \(7.5\times10^{-4}\) changes \(\mathcal{R}_{\rm all}\) from \(+26.67\) to \(-5.28\). Table~\ref{tab:1} includes the intermediate mass. Every row has its own reheating background and running Yukawa matrix, with the common \(w=0.0886066\) and fixed real rotation angles. The nonthermal \(N_2\) contribution determines the sign at all three displayed masses; the remaining sources have a positive net contribution that preserves those signs.

\begin{table*}[!htbp]
\caption{Physical mass trajectory, controlled vacuum decay parameter and source decomposition. All source contributions in a row use that mass point and its original stopping prescription. The column \(\mathcal{R}_{\rm all}-\mathcal{R}_{\rm nt2}\) measures the net contribution of the remaining three sources. The definition of \(K_1\) is Eq.~\eqref{eq:6}.}\label{tab:1}
\small
\setlength{\tabcolsep}{3.2pt}
\begin{tabularx}{\textwidth}{>{\hsize=1.4\hsize\raggedright\arraybackslash}X>{\hsize=0.9\hsize\raggedleft\arraybackslash}X>{\hsize=0.9\hsize\raggedleft\arraybackslash}X>{\hsize=0.9\hsize\raggedleft\arraybackslash}X>{\hsize=0.9\hsize\raggedleft\arraybackslash}X}
\toprule
\(M_1/M_2\) & \(K_1\) & \(\mathcal{R}_{\rm nt2}\) & \(\mathcal{R}_{\rm all}\) & \(\mathcal{R}_{\rm all}-\mathcal{R}_{\rm nt2}\) \\
\midrule
\(6.30524\times10^{-3}\) & \(1.00639\) & \(+26.04\) & \(+26.67\) & \(+0.63\) \\
\(1.5\times10^{-3}\) & \(1.00383\) & \(+1.90\) & \(+2.05\) & \(+0.15\) \\
\(7.5\times10^{-4}\) & \(1.00228\) & \(-5.35\) & \(-5.28\) & \(+0.073\) \\
\bottomrule
\end{tabularx}
\end{table*}

The two lower masses have opposite signs. Assuming continuity of the solution along this fixed-parameter path, at least one physical zero lies in

\begin{equation*}
7.5\times10^{-4}<\left({M_1\over M_2}\right)_c<1.5\times10^{-3},
\end{equation*}

corresponding approximately to \(4.98\times10^8<M_{1,c}<9.96\times10^8\ {\rm GeV}\). This interval also brackets \(\mathcal{R}_{\rm all}=1\), since its endpoint values are \(-5.28\) and \(+2.05\). The fixed-background timing comparison below provides a complementary zero.

Across the endpoint pair, the integrated norm of the nonthermal \(N_2\) injection changes by only \(3.94\times10^{-5}\) relatively, while its CP orientation and the \(N_1\) projector direction remain closely aligned. The vacuum decay parameter \(K_1\) decreases from \(1.00639\) to \(1.00228\), a relative change of about \(0.4\%\). The following washout intervention identifies the process responsible for the contrast.

Suppressing the \(N_1\) inverse-decay and \(\Delta L=1\) scattering block \(W_1\), while retaining each background and source, gives \(\mathcal{R}_{\rm all}\simeq18.45\) at both endpoints. Their difference is only \(2.18\times10^{-3}\), compared with \(31.95\) in the reference calculation. This locates the dominant contrast in the subsequent \(N_1\) washout. Carrying both thermal populations and the charge matrix from the first vacuum crossing preserves the endpoint signs.

We next resolve how the mass change moves washout through the evolving flavor charge.

\subsection{Washout timing and the fixed-history comparison}

The two lower-mass targets encounter washout on a nearly adiabatic, radiation-dominated branch. Table~\ref{tab:2} reports the small median departures from \(H/H_{\rm rad}=r_s=-d\ln T/dN=1\). Here \(H_{\rm rad}\) is the radiation-dominated reference Hubble parameter. Equal \(z_1=M_1/T\) then places a smaller mass at a later e-fold. Write \(r_M=M_1/M_2\). On the later cooling branch, equal \(M_1/T\) gives a mass-dependent delay of \(\ln(r_{M,\rm ref}/r_{M,\rm target})\). Lowering \(M_1\) therefore mainly delays washout into a more decohered charge state.

Reheating and the later radiation era play distinct roles. Table~\ref{tab:2} places the source and washout windows on the computed background.

\begin{table*}[!htbp]
\caption{Background departures from radiation scaling. Here \(\mathbf b_{{\rm nt},2}\) is the nonthermal \(N_2\) charge source. The \(N\) intervals enclose the central 5\%--95\% of the positive weights \(\|\mathbf b_{{\rm nt},2}\|_F\) and \(\|W_1\|_F\), respectively. The four rightmost columns give absolute deviations at their weighted medians; percent columns refer to \(\rho_R/\rho_{\rm tot}\) and \(r_s\).}\label{tab:2}
\small
\setlength{\tabcolsep}{3.2pt}
\begin{tabularx}{\textwidth}{>{\hsize=1.4\hsize\raggedright\arraybackslash}X>{\hsize=0.92\hsize\raggedleft\arraybackslash}X>{\hsize=0.92\hsize\raggedleft\arraybackslash}X>{\hsize=0.92\hsize\raggedleft\arraybackslash}X>{\hsize=0.92\hsize\raggedleft\arraybackslash}X>{\hsize=0.92\hsize\raggedleft\arraybackslash}X}
\toprule
Input window & \(N_{5\%}\)--\(N_{95\%}\) & \(\lvert H/H_{\rm rad}-1\rvert\) & \(\lvert1-\rho_R/\rho_{\rm tot}\rvert\), \% & \(\lvert1-r_s\rvert\), \% & \(\left|1+\frac{d\ln T}{dN}\right|\) \\
\midrule
Reference injection & 10.81--13.38 & 0.67 & 64.21 & 64.42 & 0.50 \\
Reference washout & 13.13--15.24 & \(<10^{-8}\) & \(<2\times10^{-6}\) & \(<10^{-12}\) & \(<10^{-7}\) \\
Washout, \(r_M=1.5\times10^{-3}\) & 14.65--16.68 & \(<10^{-8}\) & \(<2\times10^{-6}\) & \(<10^{-12}\) & \(<10^{-7}\) \\
Washout, \(r_M=7.5\times10^{-4}\) & 15.34--17.37 & \(<10^{-8}\) & \(<2\times10^{-6}\) & \(<10^{-12}\) & \(<10^{-7}\) \\
\bottomrule
\end{tabularx}
\end{table*}

The injection samples continuing entropy production, while the two target central washout windows are nearly adiabatic throughout. The reference washout retains an earlier nonadiabatic wing: within its central interval, \(H/H_{\rm rad}\) reaches 1.16 and the radiation fraction falls to 74.6\%.

The issue here is how to isolate the effect of washout timing when production, redistribution, and washout have extended and overlapping histories.

We isolate timing on the reference history through

\begin{equation}
\mathbf x'=
\left[A_{\rm rest}(N)+W_1(N-\delta)\right]\mathbf x+
\mathbf b_{\rm nt,2}(N),\qquad \mathbf x_i=0.
\label{eq:11}
\end{equation}

Positive \(\delta\) moves the entire \(W_1\) history to later \(N\), preserving its shape and integrated strength. The nonthermal \(N_2\) source, background and remaining operators stay fixed. This is a diagnostic intervention on one history; the physical mass points are recomputed separately.

Each displacement is determined by alignment of the normalized washout-norm profiles, without using the final abundance. The targets \(M_1/M_2=1.5\times10^{-3}\) and \(7.5\times10^{-4}\) give \(\delta_{\rm mid}=1.43627\) and \(\delta_*=2.12930\). Only the alignment profiles are normalized; the translated operator retains its amplitude.

The computed curve gives

\begin{equation}
\begin{gathered}
\mathcal{R}_{\rm nt2}(0)=+26.04,\\
\mathcal{R}_{\rm nt2}(\delta_*)=-5.25,\\
\mathcal{R}_{\rm nt2}(3)=-8.68.
\end{gathered}
\label{eq:12}
\end{equation}

The fixed-source intervention therefore reproduces a sign reversal. To evaluate its connection to the physical mass variation, Table~\ref{tab:3} compares the translated reference against both physical targets using the same single-source convention and terminal interval.

\begin{table*}[!htbp]
\caption{Comparisons of the translated history with two physical targets. Each displacement is determined from that target's operator profile. The difference is \(\mathcal{R}_{\rm nt2}^{\rm translated}-\mathcal{R}_{\rm nt2}^{\rm physical}\). The last column is the fraction of the reference-to-target change explained by translation, \([\mathcal{R}_{\rm nt2}(0)-\mathcal{R}_{\rm nt2}^{\rm translated}]/[\mathcal{R}_{\rm nt2}(0)-\mathcal{R}_{\rm nt2}^{\rm physical}]\). Differences and percentages use unrounded values, which are available in the dataset \cite{YuZhouLeptogenesisData}.}\label{tab:3}
\small
\setlength{\tabcolsep}{3.2pt}
\begin{tabularx}{\textwidth}{>{\hsize=1.4\hsize\raggedright\arraybackslash}X>{\hsize=0.92\hsize\raggedleft\arraybackslash}X>{\hsize=0.92\hsize\raggedleft\arraybackslash}X>{\hsize=0.92\hsize\raggedleft\arraybackslash}X>{\hsize=0.92\hsize\raggedleft\arraybackslash}X>{\hsize=0.92\hsize\raggedleft\arraybackslash}X}
\toprule
Target \(M_1/M_2\) & Displacement & Translated \(\mathcal{R}_{\rm nt2}\) & Physical \(\mathcal{R}_{\rm nt2}\) & Difference & Change explained \\
\midrule
\(1.5\times10^{-3}\) & \(1.4363\) & \(+2.07\) & \(+1.90\) & \(+0.170\) & \(99.3\%\) \\
\(7.5\times10^{-4}\) & \(2.1293\) & \(-5.25\) & \(-5.35\) & \(+0.103\) & \(99.7\%\) \\
\bottomrule
\end{tabularx}
\end{table*}

\begin{figure*}[t]

\centering

\includegraphics[width=\textwidth]{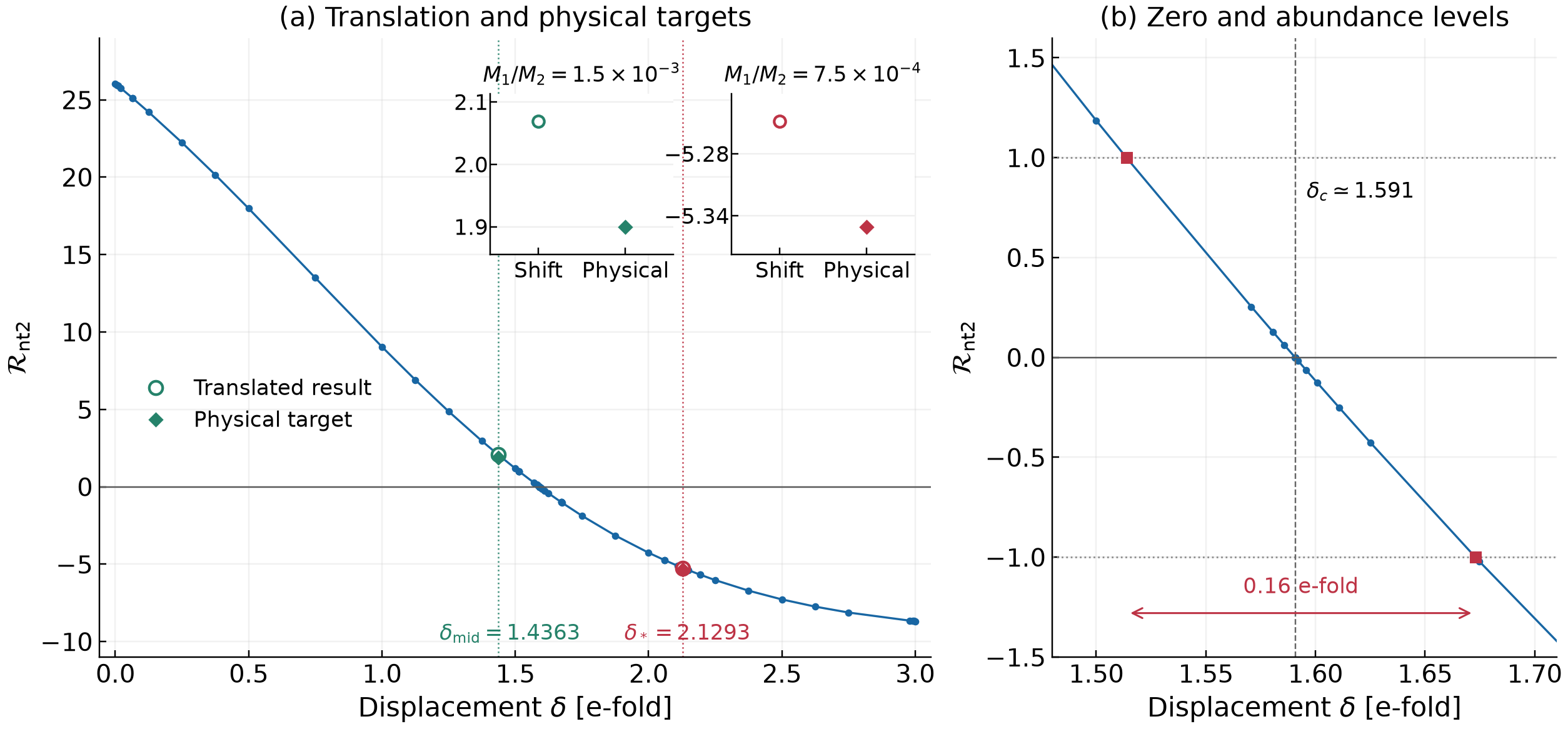}

\caption{Fixed-source timing intervention and its physical comparisons. Panel (a) shows the 62 computed values of \(\mathcal{R}_{\rm nt2}(\delta)\); connecting segments guide the eye. Open circles mark the two translated results in Table~\ref{tab:3}, and diamonds mark the physical targets at the displacement assigned by their input profiles. The small insets resolve the paired values at each alignment. Their vertical separations quantify comparison residuals. Panel (b) shows the numerical zero and the levels \(\mathcal{R}_{\rm nt2}=\pm1\). The vertical units follow Eq.~\eqref{eq:9}. All curve points share the same nonthermal \(N_2\) source, background and remaining operators.}\label{fig:2}

\end{figure*}

The two shifts determined from the washout profiles reproduce 99.3\% and 99.7\% of the abundance changes from the common reference. The physical washout histories also have 1.1\%-1.2\% smaller norm integrals and 3.9\%-4.1\% smaller widths. Their residuals in \(\mathcal{R}_{\rm nt2}\) are 0.170 and 0.103, corresponding to 8.9\% and 1.9\% of the respective terminal values. Thus timing accounts for the large physical contrast, while these remaining differences matter when predicting the smaller terminal yield. Fig.~\ref{fig:2} places both physical targets beside the fixed-source timing curve at the shifts determined from their washout profiles.

Let \(\delta_j\) be the positive e-fold displacement obtained by aligning the reference \(W_1\) norm profile with physical target \(j\), where \({\rm mid}\) and \({\rm low}\) label the two lower masses. Its difference from the logarithmic mass-ratio scaling is \(\Delta\delta_{{\rm align},j}\):

\begin{equation*}
\begin{gathered}
\delta_j=\ln\!\left(\frac{r_{M,\rm ref}}{r_{M,j}}\right)
+\Delta\delta_{{\rm align},j},\\
r_M=\frac{M_1}{M_2},
\\
j\in\{{\rm mid},{\rm low}\}.
\end{gathered}
\end{equation*}

Here \(\delta_{\rm low}=\delta_*\). The logarithms are 1.43592 and 2.12906, and the respective residuals are

\begin{equation*}
\begin{gathered}
\Delta\delta_{{\rm align},{\rm mid}}=3.54\times10^{-4},\\
\Delta\delta_{{\rm align},{\rm low}}=2.32\times10^{-4}
\quad\text{e-fold}.
\end{gathered}
\end{equation*}

Each shift follows the expected logarithmic mass scaling within \(4\times10^{-4}\) e-fold. The expression uses the measured mass ratios; the absolute \(M_2\) of each recomputed background is recorded with the inputs.

The sampled curve has one resolved zero, \(\delta_c\simeq1.591\). With \(M_1^{\rm ref}\) the mass at the reference physical point, the mass-delay relation above maps this diagnostic zero to \(M_1^{\rm ref}e^{-\delta_c}\simeq8.53\times10^8\ {\rm GeV}\), inside the physical zero bracket of Table~\ref{tab:1}. This correspondence places the fixed-history zero on the physical mass scale; the all-source physical zero is defined on the recomputed trajectory. Crossing from \(\mathcal{R}_{\rm nt2}=+1\) to \(-1\) takes about \(0.16\) e-fold with the normalization of Eq.~\eqref{eq:9}.

\section{Conversion of compensating flavor charge}\label{sec:flavor_charge}

\subsection{Matrix-dependent washout and the trace-free source}

The Hermitian matrix \(\epsilon^{(i)}\) describing charge conjugation and parity (CP) asymmetry is constructed from the same Yukawa bilinears and loop functions \cite{Y10_old03}. Its nondegenerate decay contribution is related to the vertex and self-energy CP terms of Ref. \cite{Y10_covi}. Its off-diagonal and trace-free parts are retained. Our sign convention assigns positive \(\epsilon^{(i)}\) to positive baryon-minus-lepton asymmetry; the corresponding source term in the left-handed lepton equation therefore carries \(-\epsilon^{(i)}\). The matrix orientation of a nonthermal source and its signed integral are separate inputs to subsequent propagation.

The decay losses in the population equations determine the nonthermal charge source, \(q_\ell'|_{{\rm nt},i}=-\epsilon^{(i)}\mathcal S_{{\rm nt},i}\), where \(\mathcal S_{{\rm nt},i}\) is the total decay loss per e-fold normalized to final entropy. The diagonal entries of \(\epsilon^{(i)}\) supply flavor asymmetries, and its off-diagonal entries source flavor coherence. Charge-density-matrix evolution connects the coherent and flavor-resolved limits \cite{Y10_old03,Y10_old07,Y10_covariant}.

Thermal sources are proportional to the departure from equilibrium. We use the decay CP matrix for the thermal decay and scattering source. At fixed propagation, background and nonthermal decay CP matrix, changing only this thermal source prescription leaves the nonthermal \(N_2\) contribution unchanged; it can change the all-source result.

Equation~\eqref{eq:3} applies componentwise to the charge matrices; positive semidefiniteness is not required. The gauge multiplicities are \(6,3,3,2,1\); antiparticles are not counted again in those multiplicities. In the massless free-gas approximation the charge susceptibility is \(g_XT^2/6\) for fermions and \(g_XT^2/3\) for bosons. Chemical-potential constraints and the redistribution of baryon, lepton and Higgs charges follow the framework of Refs. \cite{Y10_higgs_sphaleron,Y10_harvey}. These factors determine the relation between charge and chemical potential, and hence the normalization of a collision operator.

Hypercharge neutrality eliminates the independent Higgs-doublet charge:

\begin{equation}
q_H=\operatorname{Tr}\!\left[
\frac{-q_Q-4q_U+2q_D}{3}+q_\ell+2q_E\right].
\label{eq:7}
\end{equation}

Eliminating \(q_H\) through hypercharge neutrality retains its feedback in the chemical affinity and washout terms. All matrices, including the running inputs, are expressed in one fixed low-energy flavor basis.

Inverse decay acts on that charge through

\begin{equation}
\left.q_\ell'\right|_{D i}
=-{\gamma_{D i}\over8s_{\rm ent}HY_{\rm norm}}
\bigl(\{P_i,q_\ell\}+P_iq_H\bigr).
\label{eq:13}
\end{equation}

Here \(\gamma_{D i}=n_i^{\rm eq}d_i\), where \(n_i^{\rm eq}\) is the equilibrium number density and \(d_i\) the thermally averaged decay rate, and \(Y_{\rm norm}=15/(8\pi^2g_*)\). Here \(g_*\) is the effective number of relativistic degrees of freedom. The anticommutator with \(P_i\) makes the action depend on the flavor matrix, while the Higgs charge enters through the chemical affinity.

Taking the trace of the inverse-decay term for \(N_1\) gives the algebraic form

\begin{equation*}
\begin{aligned}
\left.(\operatorname{Tr}q_\ell)'\right|_{D1}
&=-\frac{\gamma_{D1}}{8s_{\rm ent}HY_{\rm norm}}
\left[2\operatorname{Tr}(P_1q_\ell)+q_H\right],\\
\operatorname{Tr}(P_1q_\ell)
&=\sum_\alpha(P_1)_{\alpha\alpha}q_{\ell,\alpha\alpha}
+2\sum_{\alpha<\beta}\operatorname{Re}\!
\left[(P_1)_{\alpha\beta}q_{\ell,\beta\alpha}\right].
\end{aligned}
\end{equation*}

The flavor projector \(P_1\) makes washout sensitive to the matrix encountered at that time. In particular, \(\operatorname{Tr}\{P_1,q_\ell\}=2\operatorname{Tr}(P_1q_\ell)\), so off-diagonal charge can contribute to the change of the total lepton charge. Charged-lepton interactions conserve \(B-L\), but their flavor decoherence changes the matrix on which subsequent washout acts. With compensating flavor contributions, changing this action can change the sign of the final asymmetry. Table~\ref{tab:6} evaluates these signed contributions along the complete evolving state.

The matrix also permits a decomposition that separates its total CP asymmetry from flavor information with zero initial trace. All matrices in this decomposition are evaluated at \(M_2\) in the same fixed basis:

\begin{equation*}
\epsilon_2=\operatorname{Tr}\epsilon^{(2)},\qquad
\epsilon_\parallel=\epsilon_2P_2,\qquad
\epsilon_{\rm tf}=\epsilon^{(2)}-\epsilon_2P_2.
\end{equation*}

The initially trace-free source is connected to the phantom terms discussed in Ref. \cite{Y10_old03}. Both components can contain off-diagonal entries in the charged-lepton basis. The remainder includes the trace-free pieces of both loop structures.

The two timing histories share the same \(\epsilon^{(2)}\), source envelope and background. Table~\ref{tab:4} gives the contributions \(\mathcal{R}_\parallel\) and \(\mathcal{R}_{\rm tf}\) to \(\mathcal{R}_{\rm nt2}\) after each component is propagated through the full flavor evolution.

\begin{table*}[!htbp]
\caption{Propagated components of the nonthermal \(N_2\) source. Each row uses one fixed source envelope and one propagation history. The units are those of \(\mathcal{R}_{\rm nt2}\). The decomposition is performed at \(M_2\) in the fixed flavor basis; each column is the final contribution obtained by evolving one CP source component.}\label{tab:4}
\small
\setlength{\tabcolsep}{3.2pt}
\begin{tabularx}{\textwidth}{>{\hsize=1.7\hsize\raggedright\arraybackslash}X>{\hsize=0.76667\hsize\raggedleft\arraybackslash}X>{\hsize=0.76667\hsize\raggedleft\arraybackslash}X>{\hsize=0.76667\hsize\raggedleft\arraybackslash}X}
\toprule
Washout timing & \(\mathcal{R}_\parallel\) & \(\mathcal{R}_{\rm tf}\) & Direct \(\mathcal{R}_{\rm nt2}\) \\
\midrule
Original & \(+10.65\) & \(+15.39\) & \(+26.04\) \\
Translated washout history & \(+10.56\) & \(-15.81\) & \(-5.25\) \\
\bottomrule
\end{tabularx}
\end{table*}

The projector-aligned contribution remains positive and nearly unchanged, while the initially trace-free contribution changes from \(+15.39\) to \(-15.81\), reversing the delayed total. Flavor information that initially cancels in total charge therefore determines the propagated sign.

Quantitatively, with \(\Delta\mathcal{R}_a=\mathcal{R}_a(\delta_*)-\mathcal{R}_a(0)\),

\begin{equation*}
\begin{gathered}
\Delta \mathcal{R}_\parallel=-0.083,\\
\Delta \mathcal{R}_{\rm tf}=-31.21,\\
\frac{\Delta \mathcal{R}_{\rm tf}}{\Delta \mathcal{R}_\parallel+\Delta \mathcal{R}_{\rm tf}}
\simeq99.7\%.
\end{gathered}
\end{equation*}
This attribution connects the source decomposition to the phantom terms of density-matrix leptogenesis \cite{Y10_old03} and motivates the flavor-decoherence tests below.

\subsection{Flavor decoherence and the signed charge balance}

The roles of production, washout and flavor decoherence can be seen together in the left-handed \(e\mu\) component. Selecting the dominant nonthermal \(N_2\) source, the \(N_1\) inverse-decay contribution, and the damping induced by the muon Yukawa interaction gives

\begin{equation*}
\begin{aligned}
q_{\ell,e\mu}'\supset{}&
\underbrace{-\epsilon^{(2)}_{e\mu}\mathcal S_{{\rm nt},2}}
_{\text{source term}}\\[3pt]
&-\underbrace{\frac{\gamma_{D1}}{8s_{\rm ent}HY_{\rm norm}}
\left[\{P_1,q_\ell\}_{e\mu}+(P_1)_{e\mu}q_H\right]}
_{\text{inverse decay}}\\[3pt]
&-\underbrace{\frac{\Gamma_\mu^{\rm coll}}{4H}q_{\ell,e\mu}}
_{\text{flavor decoherence}} .
\end{aligned}
\end{equation*}

The inverse-decay term is the \(e\mu\) component of Eq.~\eqref{eq:13}. Define the physical-time muon collision coefficient \(\Gamma_\mu^{\rm coll}(T)=c_E(T)|y_{\mu\mu}|^2\), where \(y_{\mu\mu}\) is the muon diagonal entry of the charged-lepton Yukawa matrix \(y_e\), and \(c_E(T)\) is the physical-time coefficient of its collision operator. Its coefficient per e-fold is \(\Gamma_\mu^{\rm coll}/H\). The symbol \(\supset\) selects these contributions within the coupled system; the calculation includes the other sources, scattering, chemical exchange and dispersion terms of the coupled system.

The charged-lepton collision operator acts on diagonal and off-diagonal components of the charge-density matrix. Its diagonal terms exchange charge between left-handed leptons, right-handed charged leptons and the Higgs. For the muon contribution, whose physical-time coefficient \(\Gamma_\mu^{\rm coll}\) was defined above, the chemical affinity is proportional to

\begin{equation}
a_\mu=\frac{q_{\ell,\mu\mu}}{2}-q_{E,\mu\mu}-\frac{q_H}{4}.
\label{eq:14}
\end{equation}

The same muon contribution damps off-diagonal \(e\mu\) and \(\mu\tau\) components at \(\Gamma_\mu^{\rm coll}/(4H)\) in the left-handed sector and \(\Gamma_\mu^{\rm coll}/(2H)\) in the right-handed sector. Setting \(a_\mu=0\) therefore does not impose the off-diagonal constraints of rapid flavor decoherence. The distinction rests on the matrix operator and charge susceptibility \cite{Y10_old06}.

To distinguish chemical equilibrium from flavor decoherence, we separately impose rapid muon chemical equilibration, rapid damping of its off-diagonal components, and their combination. Further comparisons damp only the \(e\mu\) or \(\mu\tau\) pair, in both chiral sectors. Each constraint acts consistently on the state, source and evolution equation and preserves \(B-L\).

To identify which muon-related physics permits the contrast in Eq.~\eqref{eq:12}, Table~\ref{tab:5} compares the original and displaced histories under the same rapid constraints. Rapid equilibration of the diagonal chemical affinity retains a large positive value at \(\delta=0\) and a negative value at \(\delta_*\). Rapid damping of the muon-related off-diagonal components instead makes both values negative and close to each other. The sign contrast is controlled by off-diagonal flavor dynamics.

\begin{table*}[!htbp]
\caption{Distinct constraints on the same two histories. All entries are the contribution from the fixed nonthermal \(N_2\) source in \(\mathcal{R}_{\rm nt2}\) as defined by Eq.~\eqref{eq:9}. The source, initial state and evolution equation are constrained consistently. Each pair constraint acts in both chiral sectors.}\label{tab:5}
\small
\setlength{\tabcolsep}{3.2pt}
\begin{tabularx}{\textwidth}{>{\hsize=1.7\hsize\raggedright\arraybackslash}X>{\hsize=0.65\hsize\raggedleft\arraybackslash}X>{\hsize=0.65\hsize\raggedleft\arraybackslash}X}
\toprule
Muon-sector treatment & \(\mathcal{R}_{\rm nt2}(0)\) & \(\mathcal{R}_{\rm nt2}(\delta_*)\) \\
\midrule
Finite reference rates & \(+26.04\) & \(-5.25\) \\
Rapid chemical equilibrium only & \(+26.78\) & \(-5.29\) \\
Rapid flavor decoherence only & \(-10.26\) & \(-9.74\) \\
Complete muon operator in its fast limit & \(-9.57\) & \(-9.78\) \\
Rapid damping of the \(e\mu\) pair only & \(-10.04\) & \(-9.73\) \\
Rapid damping of the \(\mu\tau\) pair only & \(+25.85\) & \(-5.26\) \\
\bottomrule
\end{tabularx}
\end{table*}

The pair comparison narrows the inference. Imposing the \(e\mu\) constraint largely removes the dependence on the two washout histories, whereas imposing the \(\mu\tau\) constraint retains it. This selects the \(e\mu\) contribution as influential within the chosen basis and parameter point. The hierarchy of charged-lepton loss rates motivates the test, while continuous sourcing maintains finite off-diagonal charge during its damping. The \(\mu\tau\) intervention has a smaller, finite effect, consistent with its nonzero signed contribution in Table~\ref{tab:6}. A small electron Yukawa coupling also leaves the muon-induced damping of \(e\mu\) present.

The corresponding all-source calculation gives \(\mathcal{R}_{\rm all}(0)=+26.67\) and \(\mathcal{R}_{\rm all}(\delta_*)=-5.24\). Applying the rapid muon-related off-diagonal constraint changes these to \(-9.65\) and \(-9.73\), respectively; rapid chemical equilibrium instead gives \(+27.41\) and \(-5.28\). Thus the sign contrast and its distinction between chemical equilibration and flavor decoherence also occur when all four sources are retained.

\begin{figure*}[t]

\centering

\includegraphics[width=0.98\textwidth]{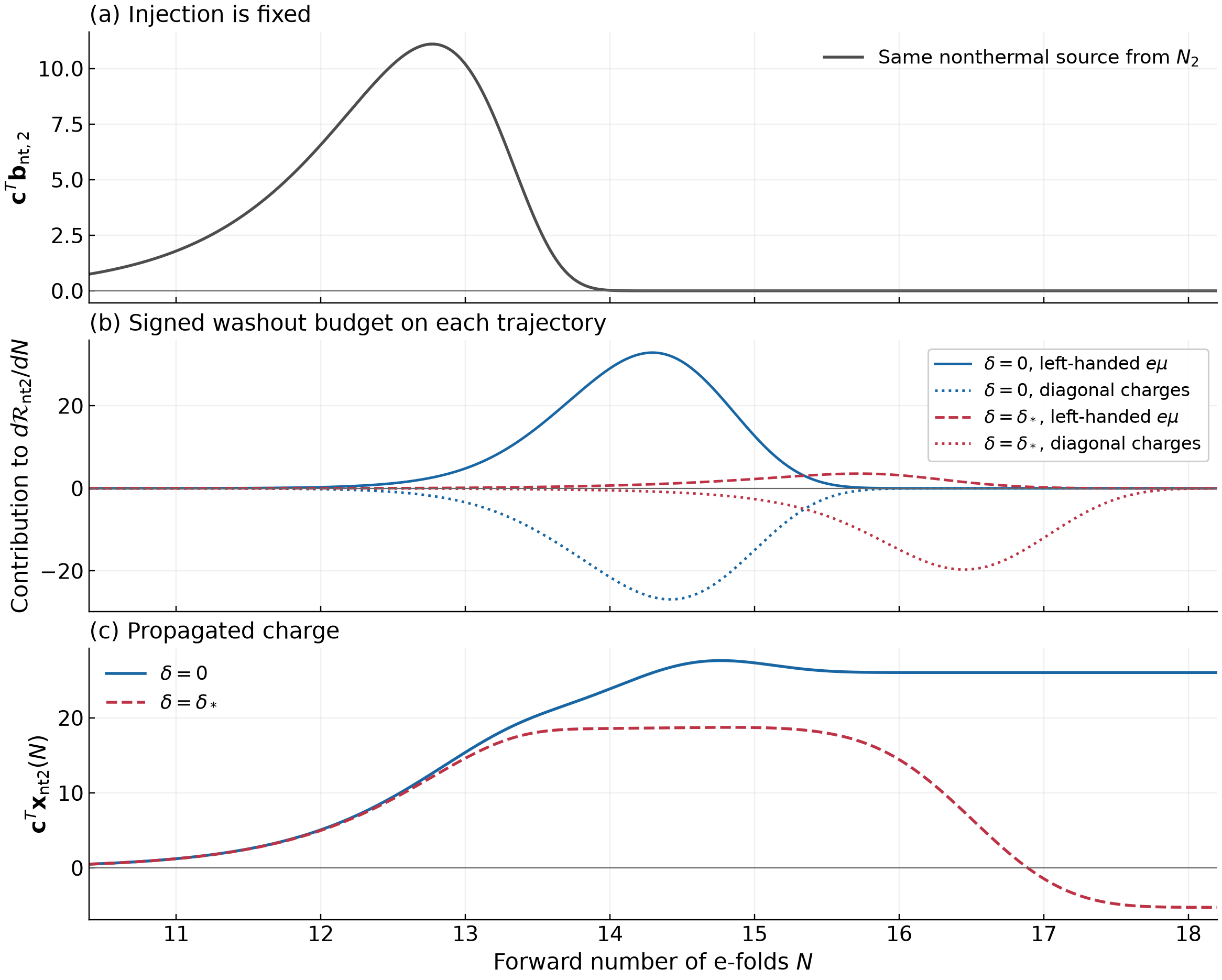}

\caption{Evolution with the original and shifted washout histories. Panel (a) shows the identical signed source injection per e-fold. Panel (b) compares the left-handed \(e\mu\) contribution to \(d\mathcal{R}_{\rm nt2}/dN\) and the sum of the quark, left-handed lepton and right-handed lepton diagonal contributions from \(W_1\). Panel (c) shows the linear functional \(\mathbf c^T\mathbf x(N)\), whose terminal value is \(\mathcal{R}_{\rm nt2}\); its intermediate values use the normalization of Eq.~\eqref{eq:9}. Blue solid and red dashed curves identify \(\delta=0\) and \(\delta=\delta_*\), respectively; dotted curves in panel (b) are diagonal contributions. All curves use the solutions of Eq.~\eqref{eq:11}.}\label{fig:3}

\end{figure*}

Chemical equilibrium and flavor decoherence have different effects on the sign contrast. Fig.~\ref{fig:3} shows the fixed injection and the changing signed evolution; Table~\ref{tab:6} gives its integrated contributions.

Table~\ref{tab:6} integrates signed contributions along each complete coupled solution. The net source injection is identical in both histories, while the positive \(N_1\) washout contribution involving left-handed \(e\mu\) charge decreases from \(47.42\) to \(6.45\). Diagonal terms change at the same time, so this decrease is only one part of the final difference. The reduction of the positive \(e\mu\) contribution is approximately \(40.96\) in \(\mathcal{R}_{\rm nt2}\), but changes elsewhere compensate part of it, leaving the actual total change \(\mathcal{R}_{\rm nt2}(\delta_*)-\mathcal{R}_{\rm nt2}(0)=-31.29\).

\begin{table*}[!htbp]
\caption{Signed contributions along the actual solutions. All entries use the nonthermal \(N_2\) source alone and the units of \(\mathcal{R}_{\rm nt2}\). Each \(W_1\) row integrates the indicated charge contribution along the complete evolving solution. The final row uses unrounded contributions.}\label{tab:6}
\small
\setlength{\tabcolsep}{3.2pt}
\begin{tabularx}{\textwidth}{>{\hsize=1.7\hsize\raggedright\arraybackslash}X>{\hsize=0.65\hsize\raggedleft\arraybackslash}X>{\hsize=0.65\hsize\raggedleft\arraybackslash}X}
\toprule
Contribution & \(\delta=0\) & \(\delta=\delta_*\) \\
\midrule
Net source injection & \(+18.47\) & \(+18.47\) \\
\(W_1\): quark diagonal charge & \(+9.07\) & \(+10.17\) \\
\(W_1\): left-handed lepton diagonal charge & \(-43.32\) & \(-25.47\) \\
\(W_1\): right-handed lepton diagonal charge & \(-5.99\) & \(-14.86\) \\
\(W_1\): left-handed \(e\mu\) charge & \(+47.42\) & \(+6.45\) \\
\(W_1\): left-handed \(e\tau\) charge & \(+0.019\) & \(+3.64\times10^{-4}\) \\
\(W_1\): left-handed \(\mu\tau\) charge & \(+0.39\) & \(+0.018\) \\
Remaining net change of baryon-minus-lepton asymmetry & \(-0.015\) & \(-0.016\) \\
Final \(\mathcal{R}_{\rm nt2}\) & \(+26.04\) & \(-5.25\) \\
\bottomrule
\end{tabularx}
\end{table*}

The source's central \(5\%\)--\(95\%\) interval in absolute integrated weight is \(N=10.81\)--\(13.38\). The corresponding interval for the left-handed \(e\mu\) washout contribution moves from \(13.11\)--\(15.05\) to \(13.85\)--\(16.55\). These intervals contain the central 90\% of each integrated contribution. The charge carrying flavor coherence is created on essentially the same schedule, while a greater part of the displaced washout acts after additional damping. Together, the source decomposition, pair constraints and signed budget provide a quantitative realization of flavor protection, phantom contributions and flavor-induced sign reversal \cite{Y10_n2,Y10_old05,Y10_old03,Y10_old08} at the two physical target masses.

A Standard Model interaction that conserves \(B-L\) can strongly affect the final asymmetry: it redistributes flavor charge before later lepton-number violation acts \cite{Y10_spectator,Y10_old04}. A single equilibration temperature compresses this history into one scale. The density-matrix evolution instead follows the competition of chemical exchange, off-diagonal damping and inverse decay \cite{Y10_zeno}. Here the two initial source components and the pair interventions resolve that competition, while the signed budget identifies its contribution to the final charge.

\subsection{Dispersion and charged-lepton collisions}

The dispersion relation couples charge to particle-plus-antiparticle sum-density deviations. Gauge interactions damp that auxiliary sector \cite{Y10_old07,Y10_flavor_review}. Its left-handed charge and sum-density equations contain

\begin{equation}
q_\ell'\supset-i[\omega_\ell,\Sigma_\ell],\qquad
\Sigma_\ell'=-i[\omega_\ell,q_\ell]-G_\ell \Sigma_\ell ,
\label{eq:15}
\end{equation}

Here \(\Sigma_\ell\) is the particle-plus-antiparticle deviation from equilibrium, normalized to the same final comoving entropy \(S_f\) as \(q_\ell\); the right-handed counterpart is \(\Sigma_E\). The matrix \(\omega_\ell\) gives flavor-dependent dispersion and \(G_\ell\) gives gauge damping per e-fold. The right-handed equation has the same structure. When the sum-density sector relaxes much faster than charge, the induced charge response scales as \((\Delta\omega)^2/G_\Sigma\), where \(\Delta\omega\) is the frequency difference between the two flavors and \(G_\Sigma\) the gauge damping coefficient. This permits strong dependence on off-diagonal collision damping together with a small dependence on the implemented dispersion relation. The numerical comparisons below test these effects directly.

The charge evolution distinguishes collisional flavor decoherence from the dispersive coupling to sum-density deviations. In the evolution equations, the effective left-handed \(e\mu\) action transmitted through the sum-density deviations is at most about \(3.41\times10^{-5}\) of the selected charged-lepton collision loss in the relevant window. The damping hierarchy explains this small effect: around \(N=14\), the physical frequency divided by \(H\) is about \(3.37\), whereas the physical gauge damping coefficient divided by \(H\) is about \(1.09\times10^6\). Their large hierarchy suppresses the charge response transmitted through the auxiliary sector.

Directly turning off the charge--sum-density cross coupling, while retaining gauge damping and the charge collision operator, changes \(\mathcal{R}_{\rm nt2}\) by \(+2.50\times10^{-3}\), \(+5.78\times10^{-4}\), and \(+3.13\times10^{-4}\) at the original history, the reference zero, and \(\delta_*\), respectively. These changes are small compared with the displacement effect. They identify the off-diagonal charge and its collision damping as the important ingredients of the present comparison.

\section{Consequences for the baryon yield}\label{sec:baryon_yield}

\subsection{Observed-abundance match}

The same mass trajectory contains a point that reproduces the observed baryon yield. Denote its mass ratio by \(r_B=(M_1/M_2)_{\widehat{\mathcal{R}}_{\rm all}=1}\). Using the late Standard Model conversion of Appendix~\ref{app:E}, \(c_{\rm eff}\simeq0.34446\), the directly recomputed point is

\begin{equation*}
\begin{gathered}
r_B\simeq1.38221\times10^{-3},\\
M_1\simeq9.18\times10^8\ {\rm GeV},\\
K_1\simeq1.00366.
\end{gathered}
\end{equation*}

Its all-source evolution starting at the first vacuum crossing gives \(Y_{B-L}\simeq2.53\times10^{-10}\) and hence \(\widehat{\mathcal{R}}_{\rm all}\simeq1\). Finite changes multiply the muon collision operator by \(\lambda_\mu\), or all charged-lepton collision operators by \(\lambda_E\), while fixing the fundamental Yukawa inputs, source, background and dispersion. These multipliers act on the charged-lepton collision coefficient while the gauge damping of the auxiliary sum-density sector stays fixed. Both multipliers include diagonal chemical exchange and off-diagonal damping. The alignment, constraint implementation and numerical verification are given in Appendices~\ref{app:C} and~\ref{app:D}.

At this fixed mass, increasing the muon collision operator by 10\% as a sensitivity test gives \(\widehat{\mathcal{R}}_{\rm all}=-0.269\). Increasing all charged-lepton collisions by the same factor gives \(-0.271\). Table~\ref{tab:7} contains these direct finite integrations with identical background, source and fundamental Yukawa inputs.

\begin{table*}[!htbp]
\caption{Direct collision tests at the late-conversion abundance match. Each row fixes the mass ratio \(r_B\) given above, evolves the state from the first vacuum crossing, includes all four sources, and uses the same \(c_{\rm eff}\). The two columns distinguish \(\mathcal{R}_{\rm all}\) of Eq.~\eqref{eq:9} from the converted baryon ratio \(\widehat{\mathcal{R}}_{\rm all}\).}\label{tab:7}
\small
\setlength{\tabcolsep}{3.2pt}
\begin{tabularx}{\textwidth}{>{\hsize=1.7\hsize\raggedright\arraybackslash}X>{\hsize=0.65\hsize\raggedleft\arraybackslash}X>{\hsize=0.65\hsize\raggedleft\arraybackslash}X}
\toprule
Collision input & \(\mathcal{R}_{\rm all}\) & \(\widehat{\mathcal{R}}_{\rm all}\) \\
\midrule
Reference & \(+0.914\) & \(1\) \\
Muon collision operator, \(\lambda_\mu=1.1\) & \(-0.246\) & \(-0.269\) \\
Charged-lepton collision operator, \(\lambda_E=1.1\) & \(-0.248\) & \(-0.271\) \\
\bottomrule
\end{tabularx}
\end{table*}

\subsection{Collision-rate response and scope}

The displacement curve also permits a local test of which collision inputs control its zero. Table~\ref{tab:C1} reports the measured derivative with respect to the muon collision multiplier. The complete muon operator and its off-diagonal part have nearly equal derivatives close to the zero and at \(\delta_*\), but differ appreciably in the original history. Chemical redistribution can therefore change the numerical response even where flavor decoherence supplies the principal distinction between signs.

\begin{equation}
\frac{d\delta_c}{d\ln\lambda_\mu}
=-\frac{\partial \mathcal{R}_{\rm nt2}/\partial\ln\lambda_\mu}
{\partial \mathcal{R}_{\rm nt2}/\partial\delta}.
\label{eq:16}
\end{equation}

This ratio measures the local movement of the zero. At the simple reference zero, \(\partial \mathcal{R}_{\rm nt2}/\partial\delta=-12.60\). Equation~\eqref{eq:16} then gives, for the complete muon operator,

\begin{equation}
\frac{d\delta_c}{d\ln\lambda_\mu}=-0.950.
\label{eq:17}
\end{equation}

Direct integrations show the root movement: reducing \(\lambda_\mu\) to \(0.93\) moves \(\delta_c\) from about \(1.591\) to \(1.660\), while increasing it to \(1.07\) moves the root to \(1.527\). The curve also changes shape. The tested finite multipliers preserve the original positive sign and the negative sign at \(\delta_*\), but give opposite signs near the reference zero. The positive final conversion factor cancels from Eq.~\eqref{eq:16}, so the root response in Eq.~\eqref{eq:17} is independent of that convention.

In Table~\ref{tab:charged_response}, \(\lambda_E\) multiplies the charged-lepton collision operator, and \(\partial_{\ln\lambda_E}\mathcal{R}_{\rm all}=\partial \mathcal{R}_{\rm all}/\partial\ln\lambda_E\) is its all-source response with the background and fundamental Yukawa inputs fixed. This differs from the muon-only response for the nonthermal \(N_2\) source in Table~\ref{tab:C1}.

\begin{table}[!htbp]
\caption{All-source response to charged-lepton collisions at the physical mass points of Table~\ref{tab:1}. Each row uses that mass point and its original stopping prescription.}\label{tab:charged_response}
\small
\setlength{\tabcolsep}{3.2pt}
\begin{tabularx}{\columnwidth}{LR}
\toprule
\(M_1/M_2\) & \(\partial_{\ln\lambda_E}\mathcal{R}_{\rm all}\) \\
\midrule
\(6.30524\times10^{-3}\) & \(-12.54\) \\
\(1.5\times10^{-3}\) & \(-13.40\) \\
\(7.5\times10^{-4}\) & \(-6.62\) \\
\bottomrule
\end{tabularx}
\end{table}

We also compare two charged-lepton prescriptions based on B{\"{o}}deker and Schr{\"{o}}der \cite{Y10_old11}. The domain hybrid uses their kernel within its published temperature interval and returns to the reference at high temperature; the full-temperature extrapolation applies it throughout. Table~\ref{tab:8} shows the timing results under the alternative kernels. Both mechanism comparisons retain the same shift determined from the washout profiles.

The alternative prescriptions already preserve opposite signs at the self-consistently recomputed physical endpoints: their all-source pairs are \((+26.68,-5.06)\) for the domain hybrid and \((+28.23,-4.34)\) for the full-temperature extrapolation. In the latter case, switching off \(W_1\) again leaves an endpoint difference of only about \(2.18\times10^{-3}\). Table~\ref{tab:8} tests timing and the \(e\mu\) discrimination under these inputs with the same \(\delta_*\) fixed by the washout profiles.

\begin{table}[!htbp]
\caption{Timing and flavor-decoherence tests with alternative charged-lepton inputs. All entries use the nonthermal \(N_2\) source, evolution from the first vacuum crossing and the common endpoint \(N=20\). The fixed translation is \(\delta_*=2.1293\). ``Fast'' means the consistently projected \(e\mu\) constraint in both chiral sectors.}\label{tab:8}
\small
\setlength{\tabcolsep}{3.2pt}
\begin{tabularx}{\columnwidth}{>{\hsize=1.4\hsize\raggedright\arraybackslash}X>{\hsize=0.9\hsize\raggedleft\arraybackslash}X>{\hsize=0.9\hsize\raggedleft\arraybackslash}X>{\hsize=0.9\hsize\raggedleft\arraybackslash}X>{\hsize=0.9\hsize\raggedleft\arraybackslash}X}
\toprule
Charged-lepton input & Finite, original & Finite, shifted & Fast, original & Fast, shifted \\
\midrule
Reference & \(+26.04\) & \(-5.25\) & \(-10.04\) & \(-9.73\) \\
Domain hybrid & \(+26.05\) & \(-5.03\) & \(-10.04\) & \(-9.72\) \\
Full-temperature extrapolation & \(+27.60\) & \(-4.30\) & \(-10.16\) & \(-9.73\) \\
\bottomrule
\end{tabularx}
\end{table}

Both alternative inputs preserve the finite-rate sign contrast. The fast constraint leaves only 1.04\% and 1.36\% of the corresponding absolute timing contrast, compared with 1.00\% for the reference. The timing contrast and its electron-muon attribution persist under both charged-lepton prescriptions.

The quantitative attribution concerns the nonthermal \(N_2\) original/translated pair, and the mass comparisons follow the stated trajectory. The finite collision multipliers measure input dependence. The kinetics use the stated thermal-bath closure, free-gas susceptibilities, thermal thresholds and scattering CP prescription. The calculations from the first vacuum crossing control the initial state within that closure.

Gauge damping uses the Born subset in Eq.~\eqref{eq:40}, multiplied by ten, with hard thermal loop and Landau-Pomeranchuk-Migdal completion outside that approximation. A threshold-dispersion screening gives an effective \(e\mu\) operator scale of about 0.084 relative to the selected collision loss; its integrated abundance effect has not been computed. The charged-lepton alternatives use the converted conductance of Ref. \cite{Y10_old11}, with a domain hybrid and a high-temperature extrapolation. A complete thermal uncertainty assessment would also require the leading soft-gauge contributions to relativistic heavy-neutrino production \cite{Y10_anisimov_rate,Y10_besak}, a rate construction across the mass regimes \cite{Y10_ghisoiu}, and collision and auxiliary terms at a common perturbative order.

At electroweak temperatures the high-scale sources, washout and entropy production have ended. In a radiation-dominated Standard Model plasma with conserved \(B-L\) and comoving entropy, evolving the late conversion gives \(c_{\rm eff}\simeq0.34446\), using leading susceptibilities with the Higgs condensate \cite{Y10_eijima_baryon_freezeout}. The coefficient converts the computed \(Y_{B-L}\) into the baryon yield; a positive overall conversion leaves the sign and zero unchanged. Finite-mass, flavor-dependent and higher-order susceptibility corrections remain outside the calculation.

\section{Conclusions}\label{sec:conclusions}

We have identified how \(N_1\) washout reverses the baryon asymmetry produced by almost unchanged nonthermal \(N_2\) injection as the heavy-neutrino mass hierarchy changes. Charged-lepton interactions evolve the injected flavor charge while inverse decay and scattering act on it. Lowering \(M_1\) delays the principal washout according to the expected logarithmic mass scaling in a nearly radiation-dominated era. Translations fixed by the washout profiles reproduce more than 99\% of the abundance change at two target masses. In the fixed-source nonthermal \(N_2\) comparison, the projector-aligned CP source gives positive contributions for both timing histories. The initially trace-free source component dominates the yield difference between them and determines the negative sign of the delayed yield, showing how flavor charge with zero initial net asymmetry can determine the final sign.

Flavor decoherence controls this conversion: later washout encounters less off-diagonal charge and a different balance of signed contributions. With the same late Standard Model conversion, the point at \(M_1/M_2\simeq1.38221\times10^{-3}\) reproduces the observed baryon yield. At this benchmark, increasing the muon collision operator by 10\% as a sensitivity test changes \(\widehat{\mathcal{R}}_{\rm all}\) from unity to \(-0.269\). The mass hierarchy therefore controls the conversion of compensating flavor charge into a net baryon asymmetry through the timing of \(N_1\) washout relative to \(e\mu\) flavor decoherence.

\begin{acknowledgments}
We thank Jin Sun for helpful discussions.
\end{acknowledgments}

\section*{Data availability}

Numerical data supporting the figures, physical mass comparisons, controlled mechanism tests, collision responses and numerical checks, together with plotting and validation code, are available in the Zenodo dataset \cite{YuZhouLeptogenesisData}. The dataset provides the numerical input sources, an article-to-data index, and documentation of the scalar, axion and conditional cosmological calculations.

\appendix

\section{Model benchmark and reheating}\label{app:A}

\subsection{Physical parameters and cosmological histories}

Motivated by the construction of Kawasaki and Yanagida \cite{Y10_old01}, we consider a real inflaton \(\phi\) with scalar potential

\begin{equation}
\begin{gathered}
x=\frac{\phi}{M\bar M_{\rm Pl}},\\
V(\phi)=v_{\rm inf}^{4}F(x)^2,\\
F(x)=1-\alpha x^2-\beta x^4-x^6-c_8x^8 .
\end{gathered}
\label{eq:1}
\end{equation}

Here \(\bar M_{\rm Pl}\) is the reduced Planck mass, \(M,\alpha,\beta,c_8\) are dimensionless, and \(v_{\rm inf}\) is an energy scale. The vacuum used for reheating is the first positive root \(x_{\rm vac}\) reached on the chosen rolling solution, with \(\phi_{\rm vac}=M\bar M_{\rm Pl}x_{\rm vac}\) and \(m_\phi^2=V_{\phi\phi}(\phi_{\rm vac})\). The main realization lies near an inflection point.

The partial \(\mathbb Z_4\) discrete symmetry assignment \(q(\phi)=2\), \(q(N_i)=1\), \(q(L_\alpha)=0\), \(q(H)=3\) permits \(\phi N_iN_i\) and \(N_iL_\alpha H\) and forbids a bare \(N_iN_i\) term. Here \(q(\cdot)\) denotes the discrete charge, while \(L_\alpha\) and \(H\) are the lepton and Higgs doublets. It also permits \(\phi^2\). This assignment supplies the mass-generation selection rule. The inflationary potential is an effective field theory with the coefficients given below; the examples use a vanishing scalar portal coupling.

For the implemented interaction linear in \(\phi\), the vacuum coupling and decay width are defined by

\begin{equation}
\begin{gathered}
y_i^\phi=\frac{M_i}{\phi_{\rm vac}},\\
\Gamma_{\phi i}=
\frac{|y_i^\phi|^2m_\phi}{8\pi}
\left(1-\frac{4M_i^2}{m_\phi^2}\right)^{3/2}
\Theta(m_\phi-2M_i).
\end{gathered}
\label{eq:2}
\end{equation}

Here \(\Theta\) is the step function that closes the two-body channel below threshold. The overall normalization follows the Majorana final-state convention used in this calculation. The dependence on \(x_{\rm vac}\) is essential when the potential away from horizon crossing changes. At fixed \(m_\phi\) and \(\phi_{\rm vac}\), differentiating the mass-dependent factor \(q_m^2(1-4q_m^2)^{3/2}\), where \(q_m=M_i/m_\phi\), gives a maximum at \(q_m=1/\sqrt{10}\). This is the maximum of the stated two-body partial width within its open channel. The backgrounds used here set \(M_2=m_\phi/\sqrt{10}\).

The main calculation uses a selected realization near an inflection point, with \(M_3/M_2=100\). Table~\ref{tab:A1} specifies its inputs and calculated background. These coefficients and mass ratios define the selected model benchmark. The mass trajectory keeps its potential and complex-rotation parameter fixed, and changes \(M_1/M_2\). The coefficient \(c_8\) is zero. The potential scale \(v_{\rm inf}\) is fixed by the scalar power spectrum amplitude \(A_s=2.1\times10^{-9}\), the adopted rounded Planck normalization at the pivot wavenumber \(k_*=0.05\ {\rm Mpc}^{-1}\) \cite{Y10_planck_params}, using the slow-roll amplitude relation.

We use the Casas--Ibarra parametrization \cite{Y10_old02}, with \(M_{\rm diag}=\operatorname{diag}(M_1,M_2,M_3)\), \(m_{\rm diag}\) the diagonal light-neutrino mass matrix and \(U_\nu\) the light-neutrino mixing matrix, in the convention

\begin{equation}
Y=\frac{1}{v_{\rm EW}}U_\nu\sqrt{m_{\rm diag}}\,
R_{\rm CI}\sqrt{M_{\rm diag}},\qquad
R_{\rm CI}R_{\rm CI}^{T}=I.
\label{eq:5}
\end{equation}

In this convention the reconstructed symmetric light mass matrix is \(-U_\nu m_{\rm diag}U_\nu^T\); its overall minus sign is absorbed in the Majorana phase convention. Changing the convention for Takagi factorization requires a corresponding change in \(U_\nu\). The general Majorana mass and mixing framework is discussed in Ref. \cite{Y10_schechter}. The condition on \(R_{\rm CI}\) is complex orthogonality, not unitarity. It preserves the light mass spectrum while allowing different heavy-state flavor projectors and charge conjugation and parity asymmetry.

\begin{table}[!htbp]
\caption{Reference model inputs and background. Here \(H_*\) is the Hubble rate at pivot-scale horizon crossing and \(N_*\) is the number of e-folds from that crossing to the end of inflation. The temperature at energy equality satisfies equality of radiation and remaining non-radiation energy. The exact input values are retained in the versioned input files.}\label{tab:A1}
\small
\setlength{\tabcolsep}{3.2pt}
\begin{tabularx}{\columnwidth}{LR}
\toprule
Quantity & Near an inflection point \\
\midrule
\(M\) & \(5\times10^{-3}\) \\
\(\alpha\) & \(2.19446\times10^{-7}\) \\
\(\beta\) & \(-5.45181\times10^{-4}\) \\
\(c_8\) & \(0\) \\
\(A_s\) & \(2.1\times10^{-9}\) \\
\(v_{\rm inf}\), GeV & \(5.4897\times10^{13}\) \\
\(M_1/M_2\) & \(6.30524\times10^{-3}\) \\
\(M_3/M_2\) & \(100\) \\
\(w\) & \(0.0886066\) \\
\(H_*\), GeV & \(7.1455\times10^8\) \\
\(m_\phi\), GeV & \(2.1005\times10^{12}\) \\
\(M_1\), GeV & \(4.1882\times10^9\) \\
\(M_2\), GeV & \(6.6424\times10^{11}\) \\
Maximum temperature, GeV & \(9.9939\times10^{10}\) \\
Temperature at energy equality, GeV & \(5.7957\times10^9\) \\
\(N_*\) & \(45.677\) \\
Scalar spectral index \(n_s\) & \(0.97504\) \\
Tensor-to-scalar ratio \(r\) & \(8.3095\times10^{-12}\) \\
Running of the scalar spectral index \(\alpha_s\) & \(+1.496\times10^{-3}\) \\
\bottomrule
\end{tabularx}
\end{table}

The light-neutrino boundary at the reference scale \(\mu_0=173.34\ {\rm GeV}\) is \((m_1,m_2,m_3)=(0,\sqrt{7.42\times10^{-5}},\sqrt{2.517\times10^{-3}})\ {\rm eV}\). The neutrino mixing matrix at this low-scale boundary uses \(\sin^2\theta_{12}=0.304\), \(\sin^2\theta_{23}=0.573\), \(\sin^2\theta_{13}=0.02219\), and Dirac phase \(195^\circ\). The mixing angles and mass-squared differences match the normal-ordering best-fit values with Super-Kamiokande atmospheric data in Table 3 of Ref. \cite{Y10_nufit}, whereas the fixed Dirac phase \(195^\circ\) matches the fit without those data. This input set is therefore a project benchmark rather than one published best-fit point. The zero lightest mass and Majorana phases are model inputs. For the reference point,

\begin{equation}
R_{\rm CI}=R_{23}(0.7)R_{12}(+iw)R_{13}(0.14),
\label{eq:18}
\end{equation}

and the Majorana phases are \((\alpha_{21},\alpha_{31})=(\pi/4,0)\). Angles in this product are in radians, and the elementary block is \(R_{12}(z)=\left(\begin{smallmatrix}\cos z&\sin z&0\\-\sin z&\cos z&0\\0&0&1\end{smallmatrix}\right)\), with analogous index replacements. The real angles remain fixed along the mass trajectory. The complex orthogonal matrix is imposed at the highest heavy threshold, \(\mu_{\rm CI}=M_3\), after alignment of the Takagi phases to the stated low-scale convention. The light mass matrix at that threshold is obtained by iterative boundary matching; it is not the low-scale diagonal mass input inserted unchanged into Eq.~\eqref{eq:5}. The numerical connection to the reported \(K_1\) is given below.

The background follows finite roll and oscillation averaging of the inflaton, with perturbative production of the two kinematically accessible heavy states \cite{Y10_turner}. The population and source equations are given below.

The renormalization group equation is applied at one loop with sequential threshold matching in the Type-I seesaw mechanism \cite{Y10_antusch_operator,Y10_old10}. Matrix inputs remain in the declared flavor basis. The physical-time coefficients multiplying the charged-lepton, up-type-quark and down-type-quark collision operator are

\begin{equation}
k_E=2Tf_E,\qquad k_u=3Tf_u,\qquad k_d=3Tf_d.
\label{eq:19}
\end{equation}

Here \(f_E,f_u,f_d\) are the dimensionless reduced kernels in the conventions of Refs. \cite{Y10_old04,Y10_old06}, and \(c_E=k_E\) in the muon collision coefficient \(\Gamma_\mu^{\rm coll}\). The kernels use running couplings at \(2\pi T\); the external Yukawa matrix is evaluated at \(T\). Division by the actual background \(H(N)\) converts each coefficient once to evolution per e-fold. The strong sphaleron transition is normalized so that the charge drift is \(12\Gamma_{\rm CS}/T^3\), where \(\Gamma_{\rm CS}\) is the Chern--Simons diffusion rate. This entails a factor \(3/2\) relative to a convention using \(8\Gamma_{\rm CS}/T^3\). The following sections specify the reduced kernels, thermal populations, source approximation and matrix washout actions used to assemble the saved operators.

The hilltop background belongs to a broader scalar and axion realization \cite{Y10_old01}. Its Peccei-Quinn symmetry history, the portal correction, axion abundance and isocurvature \cite{Y10_axion_review,Y10_visinelli}, and comparisons with cosmic microwave background and baryon acoustic oscillation data \cite{Y10_old15,Y10_old12,Y10_old13,Y10_old14}, are outside the scope of this paper. The corresponding numerical records are included in the dataset \cite{YuZhouLeptogenesisData}. No combined axion and cosmological fit has been performed at the baryon-matched mass of Table~\ref{tab:7}.

\subsection{Boltzmann equations and nonthermal injection}

The construction of momentum bins follows the distribution-function transport of Ref. \cite{Y10_hahn_boltzmann}. For a freely propagating particle, the physical momentum redshifts as \(\dot p=-Hp\), so the comoving momentum \(p_c=ap\) is constant between interactions. Let \(f_i^{\rm nt}(t,p)\) denote the nonthermal distribution of species \(N_i\), averaged over its two spin states. Its physical number density in a fixed comoving momentum bin \(g\) is

\begin{equation*}
n_{ig}(t)=\frac{2}{a^3(t)}
\int_{\text{bin }g}\frac{d^3p_c}{(2\pi)^3}
f_i^{\rm nt}\!\left(t,\frac{p_c}{a(t)}\right).
\end{equation*}

The factor two counts the spin states of a Majorana neutrino. The index \(i\) labels the heavy-neutrino species, and \(g\) labels the momentum bin. At fixed \(p_c\), the time derivative of this distribution is \(\partial_t f_i^{\rm nt}-Hp\,\partial_p f_i^{\rm nt}\), as in the distribution-function Boltzmann equation above. The prefactor \(a^{-3}\) gives the dilution term \(3Hn_{ig}\) when the equation is written for physical number density.

Inflaton decay injects particles into each bin at a rate \(J_{ig}\) per physical volume and physical time. A particle of mass \(M_i\) and energy \(E_{ig}=\sqrt{M_i^2+p_{c,g}^2/a^2}\) decays at the rate \(\Gamma_i^{\rm rest}M_i/E_{ig}\), where \(p_{c,g}\) is the representative comoving momentum of the bin and \(\Gamma_i^{\rm rest}\) is the rest-frame decay width. The factor \(M_i/E_{ig}\) accounts for time dilation. These production and loss terms give the physical-time Boltzmann equation above. We use the forward number of e-folds as the integration variable:

\begin{equation*}
N=\ln\frac{a}{a_{\rm initial}},\qquad
\frac{dN}{dt}=H,\qquad
\frac{d}{dt}=H\frac{d}{dN}.
\end{equation*}

Dividing the physical-time equation by \(H\) gives the implemented population equation,

\begin{equation*}
n_{ig}'=\frac{J_{ig}}{H}-3n_{ig}
-\frac{\Gamma_i^{\rm rest}M_i}{HE_{ig}}n_{ig}.
\end{equation*}

A prime denotes differentiation with respect to \(N\). The injection rate is \(J_{ig}=b_{ig}\,2(1+w_{\rm osc})\Gamma_{\phi i}\rho_\phi/m_\phi\), where \(w_{\rm osc}\) is the averaged inflaton equation-of-state parameter. The bin weights satisfy \(\sum_g b_{ig}=1\) and \(\sum_g b_{ig}E_{ig}=m_\phi/2\) at injection. The momentum bins retain the momentum dependence of the nonthermal population. Thermal populations are evolved alongside the charge matrices.

The decay losses in the population equations determine the source term in the charge equation. In the final-entropy convention, the nonthermal contribution is

\begin{equation}
\begin{aligned}
\mathcal S_{{\rm nt},i}(N)
&=\frac{a^3}{S_fH}\sum_g\frac{\Gamma_i^{\rm rest}M_i}{E_{ig}}n_{ig}\\
&\quad+\mathcal S_{{\rm prompt},i}+\mathcal S_{{\rm remnant},i},\\
\left.q_\ell'\right|_{{\rm nt},i}&=-\epsilon^{(i)}\mathcal S_{{\rm nt},i}.
\end{aligned}
\label{eq:20}
\end{equation}

The prompt and remnant terms include the corresponding decay losses in the same per-e-fold normalization. Thermal sources are given in Eq.~\eqref{eq:26}.

\subsection{Inflationary scale matching}

The final-entropy variables in Eq.~\eqref{eq:3} retain dilution while entropy is produced. Adding a second dilution term would double count the effect. Inflationary scale matching depends on the expansion and entropy history \cite{Y10_liddle}. A consistent pivot matching here uses the same background:

\begin{equation}
N_*=-N_{{\rm end}\to f}+\ln\frac{T_0}{T_f}
+\frac13\ln\frac{g_{s0}}{g_{sf}}
+\ln\frac{H_*}{k_*/a_0}.
\label{eq:21}
\end{equation}

In this equation \(N_*=\ln(a_{\rm end}/a_*)\), \(N_{{\rm end}\to f}=\ln(a_f/a_{\rm end})\), \(k_*=a_*H_*\), and \(g_s\) counts effective degrees of freedom in entropy. We use the conventional pivot \(k_*=0.05\,\mathrm{Mpc}^{-1}\) \cite{Y10_planck_params}, the measured present temperature \(T_0=2.7255\,\mathrm K\) \cite{Y10_fixsen_cmb_temperature}, and the standard photon-plus-neutrino entropy count \(g_{s0}=3.909\). The subscripts \(0,*,{\rm end},f\) denote the present, horizon crossing, the end of inflation, and the final entropy surface. The expansion interval \(N_{{\rm end}\to f}\) begins at the end of inflation; its origin is distinct from the coordinate used for \(N_f\) in charge evolution. Omitting the finite roll to the oscillatory regime and stopping entropy accounting at energy equality misses the contribution \(-N_{\rm roll}-\tfrac13\ln(S_f/S_{\rm eq})\). Energy equality therefore cannot automatically serve as the final matching surface.

\subsection{Renormalization scales and the vacuum decay parameter}

The usual radiation-dominated interpretation of \(K_i\) is \(K_i\simeq\Gamma_i^{\rm rest}/H_{\rm rad}(T=M_i)\), with \(\Gamma_i^{\rm rest}=[Y^\dagger Y]_{ii}(M_i)M_i/(8\pi)\) and \(H_{\rm rad}(T)=\sqrt{\pi^2g_*/90}\,T^2/\bar M_{\rm Pl}\), using \(g_*=106.75\). The quoted values use the adopted \(m_*\) in Eq.~\eqref{eq:6}. Thus \(K_i\) refers to a radiation-dominated reference expansion; the actual \(\Gamma_i/H(N)\) during reheating is a different quantity.

At a common renormalization scale, Eq.~\eqref{eq:5} and unitarity of \(U_\nu\) give

\begin{equation*}
\begin{gathered}
Y^\dagger Y={\sqrt{M_{\rm diag}}R_{\rm CI}^\dagger m_{\rm diag}
R_{\rm CI}\sqrt{M_{\rm diag}}\over v_{\rm EW}^2},\\
\widetilde m_1=(R_{\rm CI}^\dagger m_{\rm diag}R_{\rm CI})_{11}.
\end{gathered}
\end{equation*}

For fixed same-scale \(m_{\rm diag}\) and \(R_{\rm CI}\), the explicit heavy-mass dependence cancels. The tree-level vacuum decay parameter is therefore a controlled quantity of this family. In the boundary-matched calculation the high-scale light matrix and the running to each threshold must also be included. The resulting near constancy of \(K_1\) is a useful control, not evidence by itself that washout during reheating changes only in time.

The low boundary is imposed at \(173.34\ {\rm GeV}\), while the Casas--Ibarra matrix is fixed at \(M_3\simeq6.64243\times10^{13}\ {\rm GeV}\). One-loop matrix evolution with sequential tree-level threshold matching connects these boundaries \cite{Y10_old10}. Below a removed heavy threshold, lepton-number violation is represented by the dimension-five neutrino-mass operator \cite{Y10_weinberg_operator}, whose Standard Model running is given in Ref. \cite{Y10_antusch_operator}. The high-scale light matrix and heavy inputs are iterated until the low light-neutrino matrix and heavy threshold conditions agree. For the three masses of Table~\ref{tab:1} the final low-neutrino matrix residual is below \(4.2\times10^{-6}\) and the heavy-threshold residual below \(3.9\times10^{-9}\). These measure numerical boundary matching within the declared truncation.

Direct substitution of the low-scale masses and Eq.~\eqref{eq:18}, without this boundary matching, gives \(\widetilde m_1^{\rm tree}=8.40694\times10^{-4}\ {\rm eV}\) and \(K_1^{\rm tree}=0.778420\) for each mass. At the reference point, the matched light Takagi masses at \(M_3\) are instead approximately \((0,0.0111104,0.0648860)\ {\rm eV}\). Their same-scale Casas--Ibarra contraction gives \(\widetilde m_1(M_3)=1.086913\times10^{-3}\ {\rm eV}\). At the lightest threshold, the saved first Yukawa column gives

\begin{equation*}
\begin{gathered}
[Y^\dagger Y]_{11}(M_1)=1.5035623\times10^{-7},\\
\widetilde m_1(M_1)=1.0869037\times10^{-3}\ {\rm eV},\\
K_1=1.0063923.
\end{gathered}
\end{equation*}

Thus the \(29.29\%\) difference from the low-input tree diagnostic is reproduced through the declared scale matching. The unitary mixing matrix cancels in the same-scale contraction. The reference-point high-to-low-threshold change is small, while the longer intervals for the lower target masses give the finite differences in Table~\ref{tab:A2}. Repeating the three renormalization-group boundary calculations reproduces the threshold columns with relative Frobenius differences below \(5\times10^{-15}\).

\begin{table*}[!htbp]
\caption{Numerical connection between the Casas--Ibarra boundary and the vacuum decay parameter. The two effective-mass columns are in \(10^{-3}\ {\rm eV}\). The final column is computed directly from the saved Yukawa column at its own threshold, with \(v_{\rm EW}=174\ {\rm GeV}\) and \(m_*=1.08\times10^{-3}\ {\rm eV}\).}\label{tab:A2}
\small
\setlength{\tabcolsep}{3.2pt}
\begin{tabularx}{\textwidth}{>{\hsize=1.4\hsize\raggedright\arraybackslash}X>{\hsize=0.9\hsize\raggedleft\arraybackslash}X>{\hsize=0.9\hsize\raggedleft\arraybackslash}X>{\hsize=0.9\hsize\raggedleft\arraybackslash}X>{\hsize=0.9\hsize\raggedleft\arraybackslash}X}
\toprule
\(M_1/M_2\) & \(M_1\), GeV & \(\widetilde m_1(M_3)\) & \(\widetilde m_1(M_1)\) & \(K_1\) \\
\midrule
\(6.30524\times10^{-3}\) & \(4.1882140\times10^9\) & \(1.086913\) & \(1.086904\) & \(1.0063923\) \\
\(1.5\times10^{-3}\) & \(9.9636461\times10^8\) & \(1.086860\) & \(1.084133\) & \(1.0038272\) \\
\(7.5\times10^{-4}\) & \(4.9818231\times10^8\) & \(1.086834\) & \(1.082467\) & \(1.0022846\) \\
\bottomrule
\end{tabularx}
\end{table*}

\subsection{Reheating and nonthermal injection}

The background is evolved through finite roll and the subsequent oscillatory regime. Before oscillation averaging the equations are \(\ddot\phi+(3H+\Gamma_\phi)\dot\phi+V_\phi=0\) and \(\dot\rho_\phi+3H(\rho_\phi+p_\phi)=-\Gamma_\phi\dot\phi^2\). The production rate of heavy particles is \(2\Gamma_{\phi i}\dot\phi^2/m_\phi\). The factor of two counts the two Majorana particles in a decay. Oscillation averaging defines \(w_{\rm osc}=2\langle K_{\rm kin}\rangle/\rho_\phi-1\), evaluated from the potential in Eq.~\eqref{eq:1} rather than set to zero during the entire transition. The relation between the potential and the averaged equation of state follows from the oscillating-scalar analysis \cite{Y10_turner}; the decay and entropy equations implement the perturbative reheating description \cite{Y10_tmax}.

For comoving momentum bin \(g\), write \(p_{c,g}=ap_g\), \(E_{ig}=(M_i^2+p_{c,g}^2/a^2)^{1/2}\), and let \(n_{ig}\) be its physical number density. The averaged equations are

\begin{equation}
\begin{aligned}
\rho_\phi'&=-3(1+w_{\rm osc})\rho_\phi
-\frac{\Gamma_\phi}{H}(1+w_{\rm osc})\rho_\phi,\\
n_{ig}'&=\frac{J_{ig}}{H}-3n_{ig}
-\frac{\Gamma_i^{\rm rest}M_i}{HE_{ig}}n_{ig},\\
\rho_R'&=-4\rho_R+\frac1H\sum_{ig}\Gamma_i^{\rm rest}M_i n_{ig},\\
H^2&=\frac{\rho_\phi+\rho_R+\sum_{ig}E_{ig}n_{ig}}{3\bar M_{\rm Pl}^2},
\qquad \rho_R=\frac{\pi^2g_*}{30}T^4 .
\end{aligned}
\label{eq:22}
\end{equation}

Here \(J_{ig}=b_{ig}\,2(1+w_{\rm osc})\Gamma_{\phi i}\rho_\phi/m_\phi\), with \(\sum_g b_{ig}=1\) and \(\sum_g b_{ig}E_{ig}=m_\phi/2\) at injection. The calculation uses \(g_*=g_s=106.75\) in this high-temperature background. Its transitions to prompt-decay and late-remnant descriptions retain the corresponding number loss and entropy production. The resulting losses supply the source in Eq.~\eqref{eq:20}. Its prompt and remnant terms are already expressed per e-fold. Equation~\eqref{eq:20} defines \(\mathbf b_{{\rm nt},2}\) in Eq.~\eqref{eq:11} after packing the Hermitian matrices into real coordinates. The other charge sectors have no direct source in this convention. The thermal heavy-particle populations used in the charge operator are additional evolved populations; Eq.~\eqref{eq:22} specifies the background approximation, without an extra energy term for thermal heavy particles in its Friedmann equation.

\section{Kinetic and collision operators}\label{app:B}

\subsection{CP matrices and thermal populations}

For each heavy state the charge-conjugation-and-parity (CP) asymmetry matrix is evaluated with \(Y\) at the scale \(M_i\). Put \(h=Y^\dagger Y\), \(x_{ji}=M_j^2/M_i^2\), \(A_{ij}=Y_iY_j^\dagger h_{ji}\), and \(B_{ij}=Y_iY_j^\dagger h_{ij}\). The nondegenerate expression used here is \cite{Y10_old03}

\begin{equation}
\begin{aligned}
\epsilon^{(i)}={3\over32\pi h_{ii}}\sum_{j\ne i}
\Biggl[&i(A_{ij}-A_{ij}^\dagger){\mathcal X(x_{ji})\over\sqrt{x_{ji}}}\\
&+i(B_{ij}-B_{ij}^\dagger){2\over3(x_{ji}-1)}\Biggr],\\
\mathcal X(x)={2x\over3}
\Biggl[&(1+x)\ln(1+x^{-1})-{2-x\over1-x}\Biggr].
\end{aligned}
\label{eq:23}
\end{equation}

The second matrix structure is trace free. Both structures, including their off-diagonal components, are retained. The calculation uses stable large-mass-ratio evaluation of the loop function and excludes resonant heavy-state mixing.

For the bilinears in Eq.~\eqref{eq:23}, every term has an external \(Y_2\) vector on at least one side. Writing \(Q_2=I-P_2\) therefore gives \(Q_2\epsilon^{(2)}Q_2=0\). The remaining rank-one diagonal block has trace \(\epsilon_2\), so

\begin{equation*}
P_2\epsilon^{(2)}P_2=\epsilon_2P_2,\qquad
\{P_2,\epsilon_{\rm tf}\}=\epsilon_{\rm tf},\qquad
\{P_2,\epsilon_\parallel\}=2\epsilon_\parallel.
\end{equation*}

The trace-free source structure is connected to the phantom terms discussed in Ref. \cite{Y10_old03}. The anticommutator identities give its half-rate damping relative to the projector component under a pure \(P_2\) inverse-decay operator. The subsequent propagation here uses the evolution generator, including \(N_1\), spectator interactions and flavor decoherence. Both components can contain off-diagonal entries in the charged-lepton basis; the remainder combines the trace-free pieces of both loop structures.

For \(z_i=M_i/T\), the Maxwell--Boltzmann equilibrium density and the averaged vacuum-decay rate take the standard form \cite{Y10_pedestrians,Y10_old18}

\begin{equation}
\begin{gathered}
n_i^{\rm eq}={M_i^2T\over\pi^2}\mathrm K_2(z_i),\qquad
Y_i^{\rm eq}={45\over2\pi^4g_*}z_i^2\mathrm K_2(z_i),\\
d_i(T)={M_i\,[Y^\dagger Y]_{ii}(\max(T,M_i))\over8\pi}
{\mathrm K_1(z_i)\over\mathrm K_2(z_i)},\\
\gamma_{D i}=n_i^{\rm eq}d_i.
\end{gathered}
\label{eq:24}
\end{equation}

Here \(\mathrm K_\nu\) is a modified Bessel function, distinct from the vacuum decay parameter \(K_i\), and \(Y_i^{\rm eq}=n_i^{\rm eq}/s_{\rm ent}\) is an equilibrium abundance rather than a Yukawa column. The final-entropy thermal abundance \(Q_i^{\rm th}=a^3n_i^{\rm th}/S_f\) obeys

\begin{equation}
\begin{gathered}
(Q_i^{\rm th})'=-{d_i^{\rm rel}\over H}(Q_i^{\rm th}-r_sY_i^{\rm eq}),\\
d_i^{\rm rel}=d_i+
{2\gamma_{t,s}+4\gamma_{t,t}+2\gamma_{g,s}+4\gamma_{g,t}\over n_i^{\rm eq}}.
\end{gathered}
\label{eq:25}
\end{equation}

The first subscript on \(\gamma\) denotes top-quark or gauge scattering; the second denotes the direct or crossed channel. Each crossed entry is the average of its two crossed processes, so the displayed multiplicities are part of the kinetic equation.

In the reference calculations reported here, the thermal source is

\begin{equation}
\left.q_\ell'\right|_{{\rm th},i}
=-\epsilon^{(i)}{d_i^{\rm rel}\over H}
(Q_i^{\rm th}-r_sY_i^{\rm eq}).
\label{eq:26}
\end{equation}

Thus the thermal source vanishes in equilibrium. Equation~\eqref{eq:26} states the implemented approximation explicitly: the decay CP matrix multiplies the top and gauge scattering contributions as well. The hierarchical approximation motivates the top contribution; equality of gauge-scattering and decay CP asymmetries is not a general result \cite{Y10_old19}. The saved reference runs use unit multipliers for the lepton-number-violating and trace-free matrix structures and no additional hypercharge vertex correction. Separate gauge-structure variations and a hypercharge vertex subset are diagnostics in the research record, not extra terms included in the reference results. Their distinction is relevant to the collision scope in the main text.

For the principal single-source experiment, changing only the thermal gauge-scattering CP prescription in Eq.~\eqref{eq:26}, while holding \(A\), the background and the nonthermal decay CP matrix fixed, leaves \(\mathbf b_{{\rm nt},2}\) in Eq.~\eqref{eq:20} and hence \(\mathcal{R}_{\rm nt2}\) unchanged. This separation follows from the source equations. That CP prescription still affects all-source values; their presently small residual source contribution does not bound an uncomputed correction to its flavor structure. Gauge-scattering approximations in the propagation operator remain separate.

\subsubsection{Reaction densities and subtraction prescription}

The thermal scattering densities are momentum averages \cite{Y10_old18,Y10_old19}

\begin{equation}
\begin{gathered}
\gamma(T)={T\over64\pi^4}\int_{s_{\min}}^\infty
ds\,\sqrt{s}\,\mathrm K_1(\sqrt{s}/T)\,\widehat\sigma(s),\\
\widehat\sigma={2\lambda(s,m_a^2,m_b^2)\over s}\sigma(s),
\end{gathered}
\label{eq:27}
\end{equation}

where \(\lambda(x,y,z)=(x-y-z)^2-4yz\). The top-quark reduced cross sections follow Eqs. (94)--(95) of Ref. \cite{Y10_old18}, including the equal quark-mass approximation used in those expressions. Gauge scattering uses the three crossed tree amplitudes in Eqs. (3)--(5) of Ref. \cite{Y10_old19}, with massless external light states and thermal masses in the propagators. The leading thermal masses used in the regulators follow Eq. (83) of Ref. \cite{Y10_old18}:

\begin{equation}
\begin{aligned}
{m_H^2\over T^2}&={3g_2^2+g_Y^2\over16}+{y_t^2\over4}+{\lambda_H\over2},&
{m_\ell^2\over T^2}&={3g_2^2+g_Y^2\over32},\\
{m_Q^2\over T^2}&={g_3^2\over6}+{3g_2^2\over32}+{g_Y^2\over288}+{y_t^2\over16},&
{m_U^2\over T^2}&={g_3^2\over6}+{g_Y^2\over18}+{y_t^2\over8},\\
{m_W^2\over T^2}&={11g_2^2\over12},&
{m_B^2\over T^2}&={11g_Y^2\over12}.
\end{aligned}
\label{eq:28}
\end{equation}

Here \(V_H=\lambda_H(H^\dagger H)^2\). For gauge scattering let \(a_X=m_X^2/M_i^2\). The common lower threshold in \(s/M_i^2\) is \(\max[1,(1+\sqrt{a_\ell})^2,(\sqrt{a_H}+\sqrt{a_W})^2,(1+\sqrt{a_W})^2]\), which keeps the approximate crossed expressions away from thermal propagator poles. This conservative threshold prescription is part of the approximation. The reference decay term is Eq.~\eqref{eq:24}; optional thermal Higgs-decay kernels are not used in these reference histories. The resummed production rates of Refs. \cite{Y10_anisimov_rate,Y10_besak,Y10_ghisoiu} describe additional plasma physics and are not substituted for this stated threshold prescription.

For lepton-number change by two units, all three virtual heavy states and their interference are kept. Define \(\kappa_i=Y_iY_i^T/M_i\). With massless external states, the reaction tensor is represented by a real symmetric kernel \(F_{ij}(M/T)\):

\begin{equation}
\begin{gathered}
\gamma_{ab,cd}={T^6\over256\pi^5}
\sum_{ij}F_{ij}(\kappa_i)_{ab}(\kappa_j)^*_{cd},
\\
F_{ij}\longrightarrow192\quad(T\ll M_i,M_j).
\end{gathered}
\label{eq:29}
\end{equation}

The kernel follows the propagators and crossed terms of Ref. \cite{Y10_old18}. Real intermediate states are subtracted distributionally: the narrow-width \(\pi\delta(s-M_i^2)/(M_i\Gamma_i)\) contribution is removed, leaving a Hadamard finite part for the squared pole and principal-value interference terms. Removing only the imaginary part of a Breit--Wigner propagator would implement a different subtraction. The resulting charge action is

\begin{equation}
\begin{gathered}
\begin{aligned}
\left.q_\ell'\right|_{\Delta L=2}
&=-{T^6\over512\pi^5s_{\rm ent}H Y_{\rm norm}}\sum_{ij}F_{ij}\\
&\quad\times\Bigl[\tfrac12\{\kappa_i\kappa_j^\dagger,q_\ell\}
+\kappa_iq_\ell^T\kappa_j^\dagger+q_H\kappa_i\kappa_j^\dagger\Bigr],
\end{aligned}\\
Y_{\rm norm}={15\over8\pi^2g_*}.
\end{gathered}
\label{eq:30}
\end{equation}

The susceptibility normalization satisfies \(s_{\rm ent}Y_{\rm norm}=T^3/12\). This \(\Delta L=2\) term belongs to \(A_{\rm rest}\) and stays fixed when \(W_1\) is translated or suppressed.

\subsubsection{\texorpdfstring{Inverse-decay and \(\Delta L=1\) scattering washout}{Inverse-decay and Delta L = 1 scattering washout}}

The inverse-decay action is Eq.~\eqref{eq:13}. It contributes only to the left-handed lepton equation before the Higgs relation is substituted. For the scattering terms define \(\ell_i=\{P_i,q_\ell\}/4\), \(h_H=q_H/4\), \(r_i=n_i^{\rm th}/n_i^{\rm eq}\), and \(q_t=\operatorname{Tr}(P_Uq_U)/3-\operatorname{Tr}(P_Qq_Q)/6\). The projectors \(P_Q,P_U\) are obtained from the largest-singular-value directions of the running up-type Yukawa matrix. With \(\widehat\gamma=\gamma/(s_{\rm ent}HY_{\rm norm})\),

\begin{equation}
\begin{aligned}
\mathcal T_{t i}&=\widehat\gamma_{t,s}(r_i\ell_i+P_iq_t)\\
&\quad+\widehat\gamma_{t,t}[2\ell_i+(1+r_i)P_iq_t],\\
\mathcal T_{g i}&=\widehat\gamma_{g,s}(r_i\ell_i+P_ih_H)\\
&\quad+\widehat\gamma_{g,t}[2\ell_i+(1+r_i)P_ih_H],\\
q_\ell'&\supset-\mathcal T_{t i}-\mathcal T_{g i},\\
q_Q'&\supset\operatorname{Tr}(\mathcal T_{t i})P_Q,\\
q_U'&\supset-\operatorname{Tr}(\mathcal T_{t i})P_U .
\end{aligned}
\label{eq:31}
\end{equation}

For \(i=1\), Eqs.~\eqref{eq:13} and \eqref{eq:31}, with \(q_H\) eliminated using Eq.~\eqref{eq:7}, define \(W_1\). In particular, its gain coefficients contain the thermal population and equilibrium density of that history. Translating the saved \(W_1\) translates these coefficients too; recomputing them with the unshifted temperature would be a different experiment. The CP source in Eq.~\eqref{eq:20} is kept fixed.

\subsection{Standard Model charge redistribution}

For a Yukawa matrix \(y\) with right-handed and left-handed indices, let \(d=q_{\rm doublet}/g_{\rm doublet}\), \(r=q_{\rm singlet}/g_{\rm singlet}\), and \(h_H=q_H/4\). Here the charge and gauge multiplicity refer to the relevant doublet or singlet in each sector; \(q_{\rm doublet}\) is distinct from the down-quark charge matrix \(q_D\). The unit-rate actions are

\begin{equation}
\begin{aligned}
O_D&=-\tfrac12\{y^\dagger y,d\}+\sigma_Hy^\dagger yh_H+y^\dagger r y,\\
O_R&=-\tfrac12\{yy^\dagger,r\}-\sigma_Hyy^\dagger h_H+y d y^\dagger,
\end{aligned}
\label{eq:32}
\end{equation}

with \(\sigma_H=-1\) for the up sector and \(+1\) for down quarks and charged leptons. Multiplication by \(k_u/H,k_d/H,k_E/H\) gives their per-e-fold actions. Here \(k_u,k_d,k_E\) are the physical-time collision coefficients, with \(k_E=c_E\); their normalization is specified in Eq.~\eqref{eq:19}.

The matrix actions are defined by Eq.~\eqref{eq:32}. The charged-lepton reduced kernel in Eq.~\eqref{eq:19} is Eq. (102) of Ref. \cite{Y10_old06}:

\begin{equation}
\begin{aligned}
f_E={}&1.32\times10^{-3}y_t^2+3.72\times10^{-3}G\\
&+8.31\times10^{-4}G\ln(G^{-1})+4.74\times10^{-3}g_Y^2\\
&+1.67\times10^{-3}g_Y^2\ln(g_Y^{-2})+1.7\times10^{-3}G,\\
G={}&{3g_2^2+g_Y^2\over2}.
\end{aligned}
\label{eq:33}
\end{equation}

For quarks the reduced kernels follow Appendix A of Ref. \cite{Y10_old04}, including its estimated one-to-two contribution. To specify the normalization, put \(G_H=3g_2^2/2+g_Y^2/2\), \(G_Q=8g_3^2/3+3g_2^2/2+g_Y^2/18\), \(G_{u,d}=8g_3^2/3+(8/9,2/9)g_Y^2\), and \(V_{u,d}=8g_3^2/3+3g_2^2/2+(17/18,5/18)g_Y^2\). Then

\begin{equation}
\begin{aligned}
f_{u,d}={}&7.71\times10^{-4}G_H+1.32\times10^{-3}y_t^2\,\mathbf1_d\\
&+3.72\times10^{-3}(G_Q+G_{u,d})\\
&-8.31\times10^{-4}(G_Q\ln G_Q+G_{u,d}\ln G_{u,d})\\
&-7.72\times10^{-4}V_{u,d}+1.7\times10^{-3}(8g_3^2/3+3g_2^2/2),
\end{aligned}
\label{eq:34}
\end{equation}

where \(\mathbf1_d\) is one for down quarks and zero for up quarks. All couplings in Eqs.~\eqref{eq:33}--\eqref{eq:34} are evaluated at \(2\pi T\); the matrices in Eq.~\eqref{eq:32} are evaluated at \(T\).

Writing \(q=q_Q/6\), \(u=q_U/3\), \(d=q_D/3\), and \(\ell=q_\ell/2\), the weak and strong sphaleron affinities are \(a_{\rm ws}=\operatorname{Tr}(\ell+3q)/4\) and \(a_{\rm ss}=\operatorname{Tr}(2q-u-d)/6\). Their unit operators are \((O_Q,O_\ell)=(-3,-1)a_{\rm ws}I\) and \((O_Q,O_U,O_D)=(-2,1,1)a_{\rm ss}I\). The physical coefficients are \(12T c_{\rm ws}\) and \(18T c_{\rm ss}\), with

\begin{equation}
\begin{aligned}
c_{\rm ws}&=[13.7+4.49\ln(1.35/\sqrt{4\pi\alpha_2})]\alpha_2^5,\\
c_{\rm ss}&=[95.1+31.3\ln(1.41/\sqrt{4\pi\alpha_3})]\alpha_3^5,\\
\alpha_2^{-1}(2\pi T)&=0.0337^{-1}+{19/6\over2\pi}\ln{2\pi T\over91.1876\ {\rm GeV}},\\
\alpha_3^{-1}(2\pi T)&=0.1181^{-1}+{7\over2\pi}\ln{2\pi T\over91.1876\ {\rm GeV}}.
\end{aligned}
\label{eq:35}
\end{equation}

The diffusion-rate coefficients follow Eq. (8) of Ref. \cite{Y10_old04}. The weak-coupling scaling is established in Ref. \cite{Y10_arnold}, and the non-Abelian diffusion-rate calculation and gauge-group dependence are studied in Ref. \cite{Y10_moore}. Using its adopted \(N^5\) extrapolation and \(m_D^2=2g_3^2T^2\) gives the logarithmic coefficient \((8.24/2)(3/2)^5\simeq31.29\); multiplication of the unrounded coefficient by \(3.041\) gives the constant \(95.14\). Both are rounded in Eq.~\eqref{eq:35}. The \(N^5\) extrapolation is an input prescription, distinct from converting diffusion to charge drift. The values \(\alpha_2(M_Z)=0.0337\) and \(\alpha_3(M_Z)=0.1181\) are rounded weak-scale anchors retained for this separate one-loop sphaleron-rate prescription \cite{Y10_pdg_2024}; they are not the boundary-matched gauge couplings of the seesaw calculation. These expressions, the charge--sum-density coupling of Eq.~\eqref{eq:15} with the coefficients of Eqs.~\eqref{eq:39}--\eqref{eq:40}, the \(N_2\) charge washout and Eq.~\eqref{eq:30} specify \(A_{\rm rest}\). Each Standard Model charge operator conserves baryon-minus-lepton charge.

\subsubsection{Rate normalization and literature benchmarks}

For the full lepton-doublet charge, \(g_\ell=2\). At fixed vanishing right-handed and Higgs charges, Eq.~\eqref{eq:32} gives

\begin{equation*}
\begin{gathered}
\left.{dq_{\ell,\alpha\alpha}\over dt}\right|_{\rm coll}
=-{k_E\over2}|y_\alpha|^2q_{\ell,\alpha\alpha},\\
r_{\alpha L}=Tf_E|y_\alpha|^2
\quad\Longrightarrow\quad k_E=2Tf_E.
\end{gathered}
\end{equation*}

The right-handed diagonal loss is \(2r_{\alpha L}\). With only the muon Yukawa active, the left-handed \(e\mu\) loss is \(r_{\mu L}/2\), or \(\Gamma_\mu^{\rm coll}/(4H)\) per e-fold. These are different rates extracted from the same matrix operator \cite{Y10_old04,Y10_old06}.

The physical calculation and the stored radiation-dominated benchmarks in Table~\ref{tab:B1} use the independent Standard Model boundary at \(\mu=173.34\ {\rm GeV}\). Its gauge, Higgs-quartic and top-Yukawa inputs are 
\[
\begin{gathered}
(g_Y,g_2,g_3,\lambda_H,y_t)\\
=(0.35830,0.64779,1.1666,0.12604,0.93690),
\end{gathered}
\]
taken from the reference matched inputs of Ref. \cite{Y10_buttazzo_sm_boundary}. The diagonal charged-lepton Yukawa values are \((y_e,y_\mu,y_\tau)=(2.94\times10^{-6},6.07\times10^{-4},0.01021)\). The CKM angles and phase are \((\theta_{12}^q,\theta_{23}^q,\theta_{13}^q,\delta_q)=(0.227,0.0415,3.6\times10^{-3},1.139)\) radians in the standard parametrization, with \(Y_d=\operatorname{diag}(y_d,y_s,y_b)V_{\rm CKM}^{\dagger}\) and diagonal \(Y_u,Y_e\). The light-generation Yukawas and CKM parameters are rounded project boundary anchors consistent in scale and hierarchy with the measured fermion masses and mixing \cite{Y10_pdg_2024}; their exact decimals are not claimed as a single published \(\overline{\rm MS}\) fit at this scale. For the kernel of Ref. \cite{Y10_old06}, external charged-lepton Yukawa couplings run at \(T\) and the reduced kernel uses \(2\pi T\). The stored fit check of Ref. \cite{Y10_old11} instead evaluates both its kernel and external electron Yukawa at \(\pi T\). All rows use \(g_*=106.75\) and \(\bar M_{\rm Pl}=2.435\times10^{18}\ {\rm GeV}\) in \(H_{\rm rad}\).

\begin{table*}[!htbp]
\caption{Reference temperatures with explicitly matched rate definitions. Each temperature solves the indicated rate equal to \(H_{\rm rad}\); it is not an instantaneous equilibration surface in the reheating solution.}\label{tab:B1}
\small
\setlength{\tabcolsep}{3.2pt}
\begin{tabularx}{\textwidth}{>{\hsize=1.7\hsize\raggedright\arraybackslash}X>{\hsize=0.65\hsize\raggedleft\arraybackslash}X>{\hsize=0.65\hsize\raggedleft\arraybackslash}X}
\toprule
Rate definition and input & Stored implementation, GeV & Published reference, GeV \\
\midrule
Muon left-handed loss, kernel of Ref. \cite{Y10_old06} & \(1.3648\times10^9\) & \(1.30\times10^9\) \cite{Y10_old04} \\
Tau left-handed loss, kernel of Ref. \cite{Y10_old06} & \(3.6552\times10^{11}\) & \(3.70\times10^{11}\) \cite{Y10_old04} \\
Constrained electron slow mode, kernel of Ref. \cite{Y10_old11} & \(9.1776\times10^4\) & \(8.50\times10^4\) \cite{Y10_old11} \\
Same slow mode, converted kernel of Ref. \cite{Y10_old06} & \(1.0325\times10^5\) & \(8.50\times10^4\) \cite{Y10_old11} \\
\bottomrule
\end{tabularx}
\end{table*}

The first two rows test an implementation of the same rate adopted by Ref. \cite{Y10_old04}. An independent free-gas susceptibility construction reproduces the full constrained matrix of Ref. \cite{Y10_old11} to \(1.8\times10^{-15}\) in \(T^{-2}\) units, including \((\chi^{-1})_{E_eE_e}=4266/(481T^2)\). With the three \(L_\alpha-B/3\) charges and hypercharge held fixed, its slow-mode rate is \(r_{e,\rm slow}=(\chi^{-1})_{E_eE_e}\mathcal G_e\), and the conversion of our left-handed loss gives \(\mathcal G_e=r_{eL}T^2/3\). The \(21.5\%\) difference between the last row and the published temperature is retained.

\subsubsection{Alternative charged-lepton collision prescriptions}

The mechanism tests of Table~\ref{tab:8} use the same conversion of the B{\"{o}}deker--Schr{\"{o}}der reduced kernel \cite{Y10_old11} to our matrix equations. Let \(b_E=\mathcal G_e/(y_e^2T^3)\), with \(\mathcal G_e\) the electron conductance before multiplication by the inverse susceptibility. The corresponding coefficient is \(k_E^{\rm BS}=6Tb_E\); the reference is \(k_E^{\rm ref}=2Tf_E\) from Eq.~\eqref{eq:33}. We evaluate \(b_E\) with running Standard Model couplings at \(\pi T\), using the published fit

\begin{equation*}
\begin{aligned}
b_E={1\over2048\pi}\Bigl\{&y_t^2(2.82+1.48)\\
&+G_L[3.52+\ln(1/G_L)+0.776]\\
&+G_R[2.69+\ln(1/G_R)+2.03]\Bigr\},\\
G_L&=3g_2^2+g_Y^2,\qquad G_R=4g_Y^2.
\end{aligned}
\end{equation*}

The terms containing \(1.48\), \(0.776\) and \(2.03\) are the fitted collinear contribution. The published fit covers \(10^3\)--\(10^9\ {\rm GeV}\) \cite{Y10_old11}. In both transport variants, the external charged-lepton Yukawa matrices in the collision operator remain evaluated at \(T\), as in the reference calculation. This differs from the standalone electron-temperature benchmark in Table~\ref{tab:B1}, which evaluated its external electron coupling at \(\pi T\).

For the domain hybrid define \(u=\operatorname{clip}[\log_{10}(T/{\rm GeV})-8,0,1]\), where \(\operatorname{clip}[x,0,1]\) restricts \(x\) to that interval, and \(w_T=1-3u^2+2u^3\); then use

\begin{equation*}
k_E^{\rm hybrid}=k_E^{\rm ref}
\exp\!\left[w_T\ln{k_E^{\rm BS}\over k_E^{\rm ref}}\right].
\end{equation*}

This equals the alternative coefficient below \(10^8\ {\rm GeV}\), joins the reference smoothly between \(10^8\) and \(10^9\ {\rm GeV}\), and equals the reference above \(10^9\ {\rm GeV}\). Its implementation rejects temperatures below \(10^3\ {\rm GeV}\); the physical integrations reported here stop well above that lower limit. The full-temperature diagnostic instead sets \(w_T=1\) everywhere, including above \(10^9\ {\rm GeV}\). These choices change the charged-lepton collision operator through its scalar coefficient, preserving the reference susceptibilities, Yukawa matrices and other collision inputs. The two variants distinguish the published temperature domain from the extrapolated region.

\subsubsection{Spectator effects in ordered charge evolution}

The constant vector \(\mathbf c\) implements Eq.~\eqref{eq:9}, including any numerical rescaling of charge. Define \(U(N,N_s)\) by \(\partial_NU=A(N)U\) and \(U(N_s,N_s)=I\). Then

\begin{equation}
\mathbf x(N)=U(N,N_i)\mathbf x_i+
\int_{N_i}^{N}U(N,N_s)\mathbf b(N_s)\,dN_s .
\label{eq:36}
\end{equation}

The time-ordered exponential is necessary because collision operators at different \(N\) generally do not commute \cite{Y10_magnus}. Equation~\eqref{eq:36} makes the distinction precise: the source term at \(N_s\) is weighted by everything that happens after injection, rather than by a single positive survival factor common to all flavor components.

The adjoint sensitivity method \cite{Y10_adjoint} expresses this dependence without conflating an integrated source with its effect on the final observable. With \(\mathbf p'=-A^T\mathbf p\) and \(\mathbf p(N_f)=\mathbf c\), a first-order variation satisfies

\begin{equation}
\delta \mathcal{R}_s=\mathbf p(N_i)^T\delta\mathbf x_i+
\int_{N_i}^{N_f}\mathbf p^T
\bigl(\delta A\,\mathbf x+\delta\mathbf b\bigr)\,dN .
\label{eq:37}
\end{equation}

For an operator \(A_E\) that conserves baryon-minus-lepton asymmetry, \(\mathbf c^TA_E=0\), its local logarithmic input response is \(K_E=\mathbf p^TA_E\mathbf x\). Splitting \(A=A_C+A_W\), with \(\mathbf c^TA_C=0\), gives the useful identity

\begin{equation}
K_E(N)=\int_N^{N_f}
\mathbf c^TA_W(u)U(u,N)A_E(N)\mathbf x(N)\,du .
\label{eq:38}
\end{equation}

It follows by integrating \(\partial_u[\mathbf c^TU(u,N)]\), using charge conservation for \(A_C\), and multiplying by \(A_E(N)\mathbf x(N)\). Thus a charge-conserving redistribution changes the final observable only through later charge violation, under the fixed-source assumptions of this variation. The source and reaction maxima need not coincide. Finite variations also change the propagator and are evaluated by direct integration.

\subsection{Flavor decoherence and the dispersion relation}

The right-handed charged-lepton equation has the same charge--sum-density coupling structure as the left-handed equation in Eq.~\eqref{eq:15}. A commutator involving \(q_X\) alone would define a different approximation. The distinction is related to the overdamping of flavor oscillations by gauge interactions discussed in Ref. \cite{Y10_old07}.

The actual coefficients per e-fold are

\begin{equation}
\omega_\ell=\frac{T}{H}\frac{3\ln2}{8\pi^2}y_e^\dagger y_e,\qquad
\omega_E=\frac{T}{H}\frac{3\ln2}{4\pi^2}y_ey_e^\dagger ,
\label{eq:39}
\end{equation}

\begin{equation}
\begin{gathered}
G_\ell=\frac{2T}{H}
\frac{10(1.375g_Y^4+3.375g_2^4)}{8\pi^5},\\
G_E=\frac{2T}{H}\frac{10(5.5g_Y^4)}{8\pi^5}.
\end{gathered}
\label{eq:40}
\end{equation}

The gauge couplings are \(g_Y\) and \(g_2\). Equation~\eqref{eq:40} retains the stated Born annihilation subset and applies a factor of ten as a phenomenological damping prescription. The distinction between oscillation, damping and flavor equilibration in a plasma is developed in Refs. \cite{Y10_old07,Y10_flavor_review}. In the direct comparison reported in the main text, removing the associated charge--sum-density coupling changes \(\mathcal{R}_{\rm nt2}\) by at most \(2.50\times10^{-3}\) at the three tested washout timings.

When the sum-density sector relaxes much faster than the charge sector and the coefficients vary slowly on that relaxation time, let \(A_{\Sigma q}\) and \(A_{q\Sigma}\) be the charge--sum-density coupling blocks and \(G_\Sigma\) the sum-density damping block. Adiabatic elimination gives schematically

\begin{equation}
\Sigma\simeq G_\Sigma^{-1}A_{\Sigma q}q,\qquad
A_{\rm eff}^{(\Sigma)}=A_{q\Sigma}G_\Sigma^{-1}A_{\Sigma q}.
\label{eq:41}
\end{equation}

For a single pair the magnitude scales as \((\Delta\omega)^2/G_\Sigma\). A frequency exceeding the Hubble parameter is consequently insufficient to establish a dominant effect on charge. This formal distinction permits a large dependence on off-diagonal collision damping to coexist with a small dependence on the implemented dispersion relation.

\section{Controlled mechanism tests}\label{app:C}

\subsection{Translation of a fixed washout history}

For each physical target, at \(M_1/M_2=1.5\times10^{-3}\) or \(7.5\times10^{-4}\), we determine a separate displacement relative to the reference mass ratio \(6.30524\times10^{-3}\). For each input define \(f_i(N)=\|W_{1,i}(N)\|_F\) and \(I_i=\int f_i\,dN\), using the Frobenius norm with the consistent Hermitian-component metric. The alignment of the input profiles minimizes

\begin{equation}
\int\left[
\frac{f_{\rm ref}(N-d)}{I_{\rm ref}}
-\frac{f_{\rm lower\ mass}(N)}{I_{\rm lower\ mass}}
\right]^2dN,
\label{eq:42}
\end{equation}

giving \(\delta_{\rm mid}=1.43627\) and \(\delta_*=2.12930\). Only the profiles in the alignment criterion are normalized; the translated \(W_1\) in Eq.~\eqref{eq:11} keeps its original amplitude.

The direct calculation covers \(0\le\delta\le3\). Operator values before the beginning of the saved history are set to zero. The reported zero \(\delta_c\) solves \(\mathcal{R}_{\rm nt2}(\delta_c)=0\) within Eq.~\eqref{eq:11}. The interval \(\delta(\mathcal{R}_{\rm nt2}=-1)-\delta(\mathcal{R}_{\rm nt2}=+1)\) measures the displacement needed to cross the levels \(\mathcal{R}_{\rm nt2}=\pm1\) defined by Eq.~\eqref{eq:9}; it is neither a numerical root bracket nor a statistical interval.

The translated coefficients include the gain terms and thermal equilibrium densities evaluated at the original history's temperature at \(N-\delta\), while the other terms use the fixed background at \(N\). Equations~\eqref{eq:13} and \eqref{eq:31} specify the translated \(W_1\); the remaining operators in Eq.~\eqref{eq:11} are fixed.

\subsection{Projection and parameter sensitivity}

To distinguish chemical equilibrium from flavor decoherence, we separately impose a fast constraint on the muon chemical part, the muon off-diagonal part, and their sum. For a selected generator \(O\), let \(L\) span the left null space and let \(E\) embed the right null space, normalized so that

\begin{equation*}
LO=0,\qquad OE=0,\qquad LE=I.
\end{equation*}

The constrained calculation is

\begin{equation}
\mathbf x=E\mathbf z,\qquad
\mathbf z'=LAE\mathbf z+L\mathbf b,\qquad
\mathbf z_i=L\mathbf x_i .
\label{eq:43}
\end{equation}

Thus the source and initial state are projected together with the evolution operator. Each reduction is checked against \(\mathbf c^TEL=\mathbf c^T\). The reduction preserves the observable charge, including the right-handed charged-lepton contribution.

A further comparison applies rapid damping only to the \(e\mu\) pair or only to the \(\mu\tau\) pair. In each case both real components of the relevant Hermitian matrix element, its conjugate, and both chiral sectors are included. The operation is defined in the fixed flavor basis of the calculation. The signed budget in Table~\ref{tab:6} resolves chiral contributions, whereas this pair intervention acts on both chiral sectors.

Local parameter sensitivity is obtained by multiplying the selected muon collision input:

\begin{equation}
A_\lambda=A_1+(\lambda_\mu-1)\frac{\Gamma_\mu^{\rm coll}}{H}O ,
\label{eq:44}
\end{equation}

where \(A_1\) denotes the operator at unit multiplier, and \(O\) is either the complete muon charge operator or its off-diagonal part. All fundamental Yukawa parameters, the source and the background are held fixed. The zero is located again for each finite multiplier. At a simple zero, differentiating \(\mathcal{R}_{\rm nt2}(\delta_c(\lambda_\mu),\lambda_\mu)=0\) gives Eq.~\eqref{eq:16}, which measures the local movement of the zero.

\begin{table*}[t]
\caption{Local response to the muon collision input. Entries are \(\partial \mathcal{R}_{\rm nt2}/\partial\ln\lambda_\mu\) at \(\lambda_\mu=1\). The zero refers to the numerical root of the reference curve. Both columns use only the nonthermal \(N_2\) source; the columns specify distinct perturbation directions.}\label{tab:C1}
\small
\setlength{\tabcolsep}{3.2pt}
\begin{tabularx}{\textwidth}{>{\hsize=1.7\hsize\raggedright\arraybackslash}X>{\hsize=0.65\hsize\raggedleft\arraybackslash}X>{\hsize=0.65\hsize\raggedleft\arraybackslash}X}
\toprule
Location & Complete muon operator & Off-diagonal part only \\
\midrule
\(\delta=0\) & \(-12.33\) & \(-13.45\) \\
\(\delta\simeq\delta_c\) & \(-11.967\) & \(-11.964\) \\
\(\delta=\delta_*\) & \(-6.72\) & \(-6.69\) \\
\bottomrule
\end{tabularx}
\end{table*}

The lower part of the mass trajectory brackets both zero and the observed baryon abundance. The physical abundance point \(r_B\) solves

\begin{equation*}
\widehat{\mathcal{R}}_{\rm all}^{\rm ref}(r_B)=1.
\end{equation*}

The superscript denotes the reference collision prescription. Each new mass is accompanied by renormalization-group boundary matching, final-entropy pivot matching, reheating and all-source charge integration. Potential coefficients, \(w\), real rotation angles and low-energy boundary inputs stay fixed.

For the mechanism comparison, we retain the same \(\delta_*\) obtained by aligning washout profiles under each alternative charged-lepton prescription. We compare original and shifted washout with finite rates and with the fast \(e\mu\) constraint, using the nonthermal \(N_2\) source. These four calculations per prescription start at the first vacuum crossing. At the mass selected by the late-conversion abundance match, we also increase the muon and all charged-lepton collision operators by 10\% in separate sensitivity tests.

\newpage
\subsection{Source decomposition and background windows}

We evolve \(-\epsilon_\parallel\mathcal S_{{\rm nt},2}\) and \(-\epsilon_{\rm tf}\mathcal S_{{\rm nt},2}\) separately, with the identical nonthermal source envelope, background and full propagator used for the original and translated histories. The measured outputs are \(\mathcal{R}_\parallel\) and \(\mathcal{R}_{\rm tf}\), in the units of Eq.~\eqref{eq:9}; their sum is checked against the direct \(\mathcal{R}_{\rm nt2}\). We repeat this decomposition after evolving from the first vacuum crossing. This separates source structure from the later operation that removes a selected flavor pair. Numerically, \(\|\epsilon_{\rm tf}\|_F=1.918\times10^{-5}\), whereas the native lepton-number-conserving loop term has norm \(1.254\times10^{-7}\). The decomposition in Table~\ref{tab:4} includes the trace-free pieces of both loop structures.

To determine when reheating matters, we use the positive weights \(\|\mathbf b_{{\rm nt},2}\|_F\) and \(\|W_1\|_F\) with measure \(dN\), and report their median and central 5\%--95\% intervals. Within those intervals we read \(H/H_{\rm rad}\), \(\rho_R/\rho_{\rm tot}\), \(r_s\), and \(-d\ln T/dN\). For fixed \(g_s\), entropy accounting implies \(-d\ln T/dN=1-\tfrac13d\ln r_s/dN\). These input-weight windows are distinct from the signed response kernels. Approximately fixed \(M_1/T\) on a branch with \(T\propto e^{-N}\) would give \(\delta\simeq\ln(M_{1,\rm ref}/M_{1,\rm target})\); the background diagnostics test the regime needed for that interpretation.

The signed washout budget is evaluated on each actual trajectory. Let \(P_j\) extract disjoint charge coordinates, with \(\sum_jP_j=I\). These coordinate projectors differ from the heavy-neutrino flavor projectors \(P_i\). Define

\begin{equation*}
J_j(N)=\mathbf c^TW_1(N-\delta)P_j\mathbf x(N).
\end{equation*}

Here \(\mathcal{R}_{{\rm nt2},i}\) and \(\mathcal{R}_{{\rm nt2},f}\) are the initial and final values of the normalized observable for the nonthermal \(N_2\) source. Integration gives

\begin{equation}
\begin{aligned}
\mathcal{R}_{{\rm nt2},f}-\mathcal{R}_{{\rm nt2},i}
&=\int\mathbf c^T\mathbf b\,dN+\int\mathbf c^TA_{\rm rest}\mathbf x\,dN\\
&\quad+\sum_j\int J_j\,dN .
\end{aligned}
\label{eq:45}
\end{equation}

The identity measures how each term contributes along the computed solution. Removing a term changes that solution and is a different experiment.

\section{Numerical validation and boundary control}\label{app:D}

On a fixed background, the five Hermitian charge matrices and two auxiliary matrices give 63 real coordinates; evolving both thermal heavy-neutrino populations gives 65. We use NumPy \cite{Y10_numpy} and SciPy \cite{Y10_scipy} with a backward differentiation formula for stiff equations \cite{Y10_hairer}. The fixed-source runs start at the maximum-temperature surface. The physical abundance match and tests with alternative collision prescriptions evolve both populations and the full charge state from the first vacuum crossing. The charge evolution remains linear: the mass comparisons satisfy \(\max|\mu_{\rm chem}/T|<2.1\times10^{-4}\). For the physical mass points in Table~\ref{tab:1}, the source sum closes against the directly evolved total to within \(1.7\times10^{-6}\).

The sum of the two source contributions closes to within \(2.8\times10^{-6}\). Repeating the decomposition with evolution from the first vacuum crossing changes the displayed contributions by less than \(3\times10^{-6}\) and gives the same attribution. The constrained values are supported by finite-rate checks. The maximum closure residual across the 12 budget analyses is \(2.71\times10^{-6}\), much smaller than the signed contributions in Table~\ref{tab:6}. The signs at the numerical bracket endpoints recorded in the dataset \cite{YuZhouLeptogenesisData} persist on a finer operator grid. Backward differentiation and Radau agree on the zero within the bracket.

The maximum-temperature reset can omit both thermal heavy-neutrino populations and trace-free flavor charge. We therefore carry the two number densities and all 63 charge/auxiliary coordinates from the first vacuum crossing through the temperature maximum. The thermal number solution is continued together with the charge state when constructing subsequent thermal sources and washout. Carrying both thermal populations and the charge matrix from the first vacuum crossing preserves the endpoint signs, with changes below \(10^{-5}\) in \(\mathcal{R}_{\rm all}\). Detailed solver settings, convergence comparisons and initial-history records are provided in the dataset \cite{YuZhouLeptogenesisData}.

\newpage
\section{Late baryon conversion}\label{app:E}

At electroweak temperatures the high-scale sources, washout and entropy production have ended. We use the leading Standard Model susceptibility and relaxation relations \cite{Y10_eijima_baryon_freezeout}

\begin{equation*}
\begin{gathered}
\chi(T)=\frac{4(27x^2+77)}{333x^2+869},\\
\Gamma_B=9\frac{869+333x^2}{792+306x^2}\frac{\Gamma_{\rm diff}}{T^3},
\\
x=\frac{v(T)}{T}.
\end{gathered}
\end{equation*}

Here \(\chi(T)\) is the equilibrium ratio \(Y_B/Y_{B-L}\), \(\Gamma_B\) is the baryon relaxation rate, \(\Gamma_{\rm diff}\) is the sphaleron diffusion rate, and \(v(T)\) is the Higgs condensate. All three generations and fast Standard Model chemical constraints enter these leading massless susceptibilities. The condensate is retained; finite particle-mass and higher-order susceptibility corrections are omitted. With \(b=Y_B/Y_{B-L}\), conserved comoving entropy and \(\dot T=-HT\), the scalar evolution is

\begin{equation*}
\frac{db}{dT}=\frac{\Gamma_B}{HT}[b-\chi(T)].
\end{equation*}

We use \(v/T=3\sqrt{1-T/(159.5\ {\rm GeV})}\) \cite{Y10_hong_sphaleron_decoupling} only in the electroweak interval, and \(\Gamma_{\rm diff}/T^4=\exp[-147.7+0.83T/{\rm GeV}]\) \cite{Y10_donofrio}. Radiation domination with \(g_*=g_s=106.75\) fixes \(H\). Starting with \(b=\chi\) at 150 GeV and integrating to 125 GeV gives \(c_{\rm eff}=b_f\simeq0.34446\). Instantaneous freezing at 131.7 GeV would give 0.34313 in the same susceptibility approximation.

Any positive final coefficient rescales the abundance but leaves its sign, a zero \(\delta_c\), and the derivative ratio \(- (\partial \mathcal{R}_{\rm nt2}/\partial\ln\lambda_\mu)/(\partial \mathcal{R}_{\rm nt2}/\partial\delta)\) unchanged. Abundance values, signed budgets, abundance derivatives and intervals defined by \(\mathcal{R}_{\rm nt2}=\pm1\) do depend on the unit. For example, the symmetric-phase equilibrium ratio \(28/79\) \cite{Y10_harvey} would multiply the displayed numbers by \(1.12518\). It is distinct from an evolution through electroweak freeze-out. Equation~\eqref{eq:9} preserves the transport results while making the physical abundance comparison explicit. For comparison, Ref. \cite{Y10_babu_minimal_so10} uses 0.315 together with a top-only free-gas formula. We use the dynamical calculation above for the late conversion and retain 0.315 as the normalization of Eq.~\eqref{eq:9}.

The numerical endpoint at 130 GeV gives 0.344469, only \(5.27\times10^{-6}\) above the adopted result. Thus extending the rate fit below its quoted 130 GeV range controls a negligible tail. Changing the crossover parameter to 158 or 161 GeV gives 0.34485 or 0.34410; multiplying the diffusion input by one half or two gives 0.34471 or 0.34422. These numerical and input comparisons accompany the susceptibility approximation stated above.

\end{document}